\documentclass[twocolumn, trackchanges, resetfootnote]{aastex7}
\hypersetup{linkcolor=red, citecolor=blue, filecolor=magenta, urlcolor=cyan}

\usepackage{soul}

\renewcommand{\figurename}{Figure\,}
\renewcommand{\tablename}{Table\,}
\newcommand{\sectionname}{Section\,}
\renewcommand{\appendixname}{Appendix\,}
\def\figref#1{\figurename\,\ref{#1}}

\def\secref#1{\sectionname\,\ref{#1}}
\def\appref#1{\appendixname\,\ref{#1}}
\def\eqref#1{(\ref{#1})}

\usepackage{CJKutf8}

\shorttitle{CCs Bidirectional Expansion in SMC by \textit{Gaia}}
\shortauthors{Nakano et al.}
\graphicspath{{./}{figures/}}

\begin{document}

\title{Dual Directional Expansion of Classical Cepheids in the Small Magellanic Cloud \\ Revealed by \textit{Gaia} DR3}

\correspondingauthor{Satoya Nakano}
\email{nasatoya@a.phys.nagoya-u.ac.jp}

\author[0009-0001-8036-7296]{Satoya Nakano (中野 覚矢)}
\affiliation{Nagoya University, Nagoya, Aichi, 464-8602 Japan}
\email{nasatoya@a.phys.nagoya-u.ac.jp}

\author[0000-0002-1411-5410]{Kengo Tachihara (立原 研悟)}
\affiliation{Nagoya University, Nagoya, Aichi, 464-8602 Japan}
\email{k.tachihara@a.phys.nagoya-u.ac.jp}


\begin{abstract}

We present the three-dimensional kinematics of classical Cepheids (CCs) in the Small Magellanic Cloud (SMC) using \textit{Gaia} DR3 data. 
By cross-matching the CCs obtained from the fourth phase of the Optical Gravitational Lensing Experiment with \textit{Gaia} DR3, we obtain distances and proper motions (PMs) for 4,236 CCs.
Among them, 91 stars with available radial velocities (RVs) enable the construction of accurate relationship between distance and RV.
Furthermore, we calculate the internal PMs of the CCs, providing the internal dynamics of the SMC while removing the distance projection effects.
The CCs exhibit a northeast-southwest distance gradient, with internal PMs pointing northeast for nearer stars and southwest for more distant stars, indicating a northeast-southwest elongation. 
The RVs of the CCs show a northwest-southeast gradient, consistent with RVs of other stellar populations and suggesting that the SMC’s northwest-southeast elongation results from interactions with the Large Magellanic Cloud.
The distances and RVs of the CCs are nearly uncorrelated, implying that the two elongations arise from distinct causes.

\end{abstract}

\keywords{\uat{Cepheid distance}{217} --- \uat{Small Magellanic Cloud}{1468} --- \uat{Galaxy interaction}{600} --- \uat{Proper motions}{1295} --- \uat{Radial velocity}{1332} --- \uat{Gaia}{2360}}


\begin{CJK}{UTF8}{min}

\section{Introduction}
\label{sec:1}
The Small Magellanic Cloud (SMC), located at $\sim$60~kpc \citep{2015AJ....149..179D}, together with the Large Magellanic Cloud (LMC) at $\sim$50~kpc \citep{2014AJ....147..122D}, represents the nearest interacting galaxy system to the Milky Way.

Understanding the internal kinematics of the two Magellanic galaxies is crucial for reconstructing their interaction history with the Milky Way and their evolutionary pathways.
The kinematic signatures of the SMC and LMC are likely imprinted in the motions of the Magellanic Stream \citep{1974ApJ...190..291M} and Bridge \citep{1963AuJPh..16..570H, 1985Natur.318..160I}, thought to have formed from material stripped from these galaxies. Accurately reproducing these features is essential for constraining their orbital histories (e.g., \citealt{2007ApJ...668..949B, 2012MNRAS.421.2109B, 2012ApJ...750...36D, 2015ApJ...815...77S}). Additionally, the rotation curves of the galaxies provide key constraints on their masses, a fundamental parameter in gravitational interactions.

The mass of the SMC has traditionally been estimated using its rotation curve derived from H\,\textsc{i} emission line observations \citep{2004ApJ...604..176S, 2019MNRAS.483..392D}.
However, recent studies based on proper motions (PMs) and radial velocities (RVs) of various stellar populations have highlighed discrepancies between the three-dimensional stellar kinematics and the H\,\textsc{i} rotation \citep{2019ApJ...887..267M, 2020MNRAS.495...98D, 2021MNRAS.502.2859N, 2025arXiv250212251N}.
These findings suggest that the velocity gradient observed in H\,\textsc{i} may not reflect true galactic rotation but rather result from tidal interactions.

Interpreting the kinematics of the SMC is complicated by its significant line-of-sight depth. For example, classical Cepheids (CCs; \citealp{2017MNRAS.472..808R}) and RR Lyrae stars \citep{2017AcA....67....1J} suggest tidal stretching, with a line-of-sight extent of $\sim$20--30~kpc. Thus, the precise locations of H\,\textsc{i} and the young stars tracing it within the 50 to 80~kpc range remain unclear. The multiple line-of-sight velocity components of H\,\textsc{i} observed in the SMC may result from overlapping structures at different distances (e.g., \citealp{2007MNRAS.381L..11M}). Furthermore, since the apparent PM of objects depends on their distance, accurate determining the direction and magnitude of internal PMs requires precise stellar distance measurements.

Discussing the internal PMs of CCs, which have accurately determined distances, is important for understanding the dynamics of the SMC. \citet{2021MNRAS.502.2859N} emphasized the impact of distance on estimating internal PMs and cross-matched the CCs from \citet{2017MNRAS.472..808R} with \textit{Gaia} Data Release 2 (DR2) to examine their motion, accounting for distance.
Their results showed that nearer CCs are moving northeast relative to more distant ones, suggesting that stars in the vicinity are being stripped from the SMC. However, due to the limited RV data for the CCs in \textit{Gaia} DR2, \citet{2021MNRAS.502.2859N} had to rely on RVs from younger stellar populations \citep{2008MNRAS.386..826E} for their analysis.
Since the spatial distribution of CCs lacks the clear Wing structure seen in young stars or the interstellar medium (ISM), it remains uncertain whether the RVs of CCs align with those of young stars.

The relationship between the distances and RVs of stars in the SMC is also crucial for comparisons with theoretical simulations. \citet{2012ApJ...750...36D} modeled the SMC as a spheroid and a rotating disk, simulating its interactions with the LMC and the Milky Way. Their results showed that the disk reproduced a northeast-southwest line-of-sight velocity gradient, consistent with H\,\textsc{i} observations, while the spheroid produced a northwest-southeast gradient. Both components displayed line-of-sight velocity amplitudes of $\sim$100--200~km~s$^{-1}$. The spheroid’s velocity gradient was interpreted as a result of tidal forces from the LMC, located to the southeast, and it differed in angle from the H\,\textsc{i} velocity gradient. Furthermore, they demonstrated that the disk component forms a tidal arm, the Counter-Bridge, alongside the Magellanic Bridge. The Counter-Bridge extends behind the SMC in an arc toward the LMC, with an observational signature of increasing RVs with distance, spanning $\sim$100--200~km~s$^{-1}$—similar to the SMC’s main body. \citet{2012ApJ...750...36D} emphasized that correlating distances with RVs is key to validating the Counter-Bridge, as it overlaps with the SMC's line-of-sight.

In this study, we present the first internal PM map of the SMC that accounts for its large line-of-sight depth by cross-matching the CCs from \citet{2017MNRAS.472..808R} with \textit{Gaia} DR3 \citep{2016A&A...595A...1G,2023A&A...674A...1G}. As \textit{Gaia} DR3 provides RV measurements for sufficiently bright ($G\lesssim15$~mag) stars, a key feature of this work is the accurate relationship between distances and RVs for the SMC’s bright CCs. Our \textit{Gaia} DR3 sample includes 91 CCs with available RVs, a significant increase from the 8 CCs from \textit{Gaia} DR2. Furthermore, we identify young massive Cepheids as gas tracers by selecting massive star candidates ($M_\mathrm{ini}\geq8M_\odot$, $Age\lesssim\mathrm{50.0~Myr}$) from \citet{2025arXiv250212251N}, using the \textit{Gaia} color-magnitude diagram. \citet{2012ApJ...750...36D} suggests that the Counter-Bridge may only be observable in young stellar populations, similar to the Magellanic Bridge. Therefore, identifying particularly young stars among the CCs is crucial.
We describe the sample and internal PM calculation method in \secref{sec:sample}, present the results in \secref{sec:result}, and discuss the relationship between the distances and RVs of CCs and their spatial distribution compared to the velocity components of H\,\textsc{i} in \secref{sec:discussion}.

For this paper, we adopt a representative distance to the SMC of $D_\mathrm{SMC}=62$~kpc \citep{2014ApJ...780...59G,2015AJ....149..179D}.

\section{Sample of Classical Cepheids}
\label{sec:sample}

\subsection{Distance Calculation for CCs}
\label{sec:distance}
For the catalog of Cepheids in the SMC, we downloaded the data of CCs associated with the SMC from the database of the fourth phase of the Optical Gravitational Lensing Experiment\footnote{The OGLE Cepheids catalog can be downloaded from \url{https://ogledb.astrouw.edu.pl/~ogle/OCVS/ceph_query.php}.} (OGLE-IV survey; \citealp{2015AcA....65....1U, 2015AcA....65..297S}). 
We use the celestial coordinates and $V$ magnitudes from the OGLE-IV data. The CCs are then matched with the pulsation modes and periods ($P$) of OGLE CCs listed in Table~2 of \citet{2017MNRAS.472..808R}, along with the $K\mathrm{s}$ magnitudes from the VISTA near-infrared YJKs survey of the Magellanic System (VMC survey; \citealp{2011A&A...527A.116C}). In total, 4793 CCs are successfully matched.

To determine the distances of these CCs, we adopt the method suggested by \citet{2016ApJS..224...21R, 2017MNRAS.472..808R}. They established a reddening-free distance estimation method by deriving a more precise period–Wesenheit ($PW$) relation using near-infrared observational data, rather than the conventional period–luminosity ($PL$) relation. We adopt the following $PW$ relation, using the Wesenheit index $W(V,\,K\mathrm{s})=K\mathrm{s}-0.13(V-K\mathrm{s})$:
\begin{eqnarray}\label{1}
W(V,\,K\mathrm{s}) = \alpha\mathrm{log}P + \beta ,
\end{eqnarray}
where $\alpha$ and $\beta$ are constants that depend on the pulsation modes, as determined by \citet{2017MNRAS.472..808R}. The specific values of $\alpha$ and $\beta$ are listed in Table~3 of \citet{2017MNRAS.472..808R}. Since values of $\alpha$ and $\beta$ for mixed modes are not provided, we exclude 290 mixed-mode CCs from our sample.

The distance of the CCs, $D_\mathrm{star}$, is calculated using the following equation:
\begin{eqnarray}\label{2}
D_\mathrm{star}(V,\,K\mathrm{s})=D_\mathrm{SMC}10^{\Delta W(V,\,K\mathrm{s})},
\end{eqnarray}
where $\Delta W(V,\,K\mathrm{s}) = W(V,\,K\mathrm{s})-(\alpha\mathrm{log}P + \beta)$.
For each pulsation mode, we iteratively apply a 3.5$\sigma$-clipping procedure to $\Delta W(V,\,K\mathrm{s})$ until convergence, excluding 257 CCs from the sample in the process.
\citet{2017MNRAS.472..808R} adopted a distance to the SMC of $63~\mathrm{kpc}$, whereas we use $D_\mathrm{SMC}=62~\mathrm{kpc}$ \citep{2015AJ....149..179D}, consistent with the $62.1~\mathrm{kpc}$ \citep{2014ApJ...780...59G} used by \citet{2021MNRAS.502.2859N}.

Through the above calculations, we obtain $D_\mathrm{star}$ for a total of 4,246 CCs.
The mean $D_\mathrm{star}$ of the CCs we calculated is 62.2~kpc. The range of $D_\mathrm{star}$ for the CCs extends from $\sim$50--80~kpc, which is consistent with the results of \citet{2017MNRAS.472..808R}.

\begin{figure}
 \centering
 \includegraphics[width=8.5cm,clip]{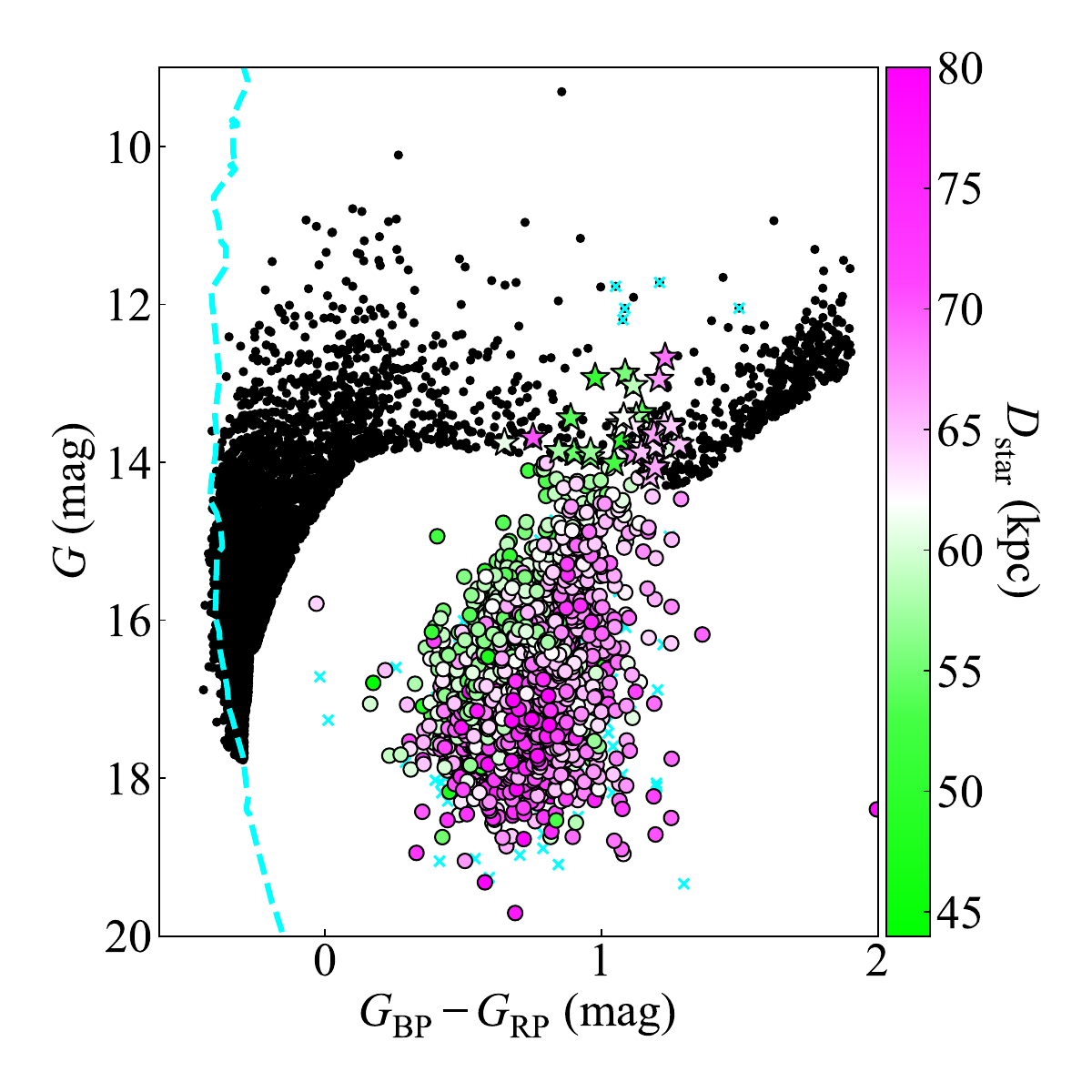}
 \caption{
 The ($G_\mathrm{BP}-G_\mathrm{RP}$, $G$) CMD of the CC samples.
 The colored circles and star symbols represent the CC samples, with their colors indicating the respective values of $D_{\mathrm{star}}$.
 The black circles represent massive star candidates ($M_{\mathrm{ini}} \geq 8M_\odot$, $\mathrm{Age} \lesssim 50.0~\mathrm{Myr}$; \citealt{2025arXiv250212251N}).
 The colored star symbols represent massive Cepheids, corresponding to 38 massive star candidates among the CC samples.
 The cyan crosses represent 543 objects that have counterparts in \textit{Gaia} DR3 but were excluded from the CC samples due to either the 3.5$\sigma$-clipping applied to $\Delta W(V,\,K\mathrm{s})$ or the lack of available PM data.
 The cyan dashed line represents an approximate main sequence constructed using isochrones of the PARSEC (version 1.2S; \citealp{2012MNRAS.427..127B, 2014MNRAS.444.2525C, 2015MNRAS.452.1068C, 2014MNRAS.445.4287T, 2017ApJ...835...77M, 2019MNRAS.485.5666P, 2020MNRAS.498.3283P}).
 The metallicity of the main sequence is assumed to be $Z = 0.1Z_\odot$, with $Z_\odot = 0.0152$.
 Additionally, a reddening correction of $A_V = 0.15~\mathrm{mag}$ due to the Milky Way foreground is applied to the main sequence using \texttt{dustapprox} \citep{Fouesneau_dustapprox_2022}.
}
\label{CEP_CMD}
\end{figure}

\subsection{Cross-matching with \textit{Gaia} DR3}
\label{sec:crossmatch}
We cross-match the 4,246 CCs obtained in \secref{sec:distance} with \textit{Gaia} DR3.
A cross-match radius of 0.5~arcsec is adopted, and \textit{Gaia} DR3 counterparts are found for all CCs.
However, eight CCs have two \textit{Gaia} DR3 sources within the radius. To ensure reliability, we exclude them from our sample\footnote{The OGLE-IDs of the excluded objects are \texttt{OGLE-SMC-CEP-1060}, \texttt{1617}, \texttt{1623}, \texttt{2683}, \texttt{2752}, \texttt{2863}, \texttt{3593}, and \texttt{4101}.}.
By further removing sources without available PM measurements, our final CC sample consists of 4,236 objects.
Hereafter, we refer to these 4,236 CCs simply as the CC samples.
A complete list of the CC samples, along with various physical parameters, is provided in \appref{sec:catalog}.

Below, the PM in the RA direction, corrected for coordinate distortion by multiplying with $\mathrm{cos}\,\delta\,$, is denoted as $\mu_\alpha$, while the PM in the DEC direction is denoted as $\mu_\delta$.
For the CC samples, the mean errors of $\mu_\alpha$ and $\mu_\delta$ are $(0.075,\,0.069)~\mathrm{mas~yr^{-1}}$.
The tangential velocities in units of $\mathrm{km~s^{-1}}$ are obtained by multiplying the PMs by $4.7438 \times D_\mathrm{star}~\mathrm{(kpc)}$.
For the CC samples, the mean errors of the tangential velocities are $(23.3,\,23.0)~\mathrm{km~s^{-1}}$, assuming a uniform uncertainty of $1.6~\mathrm{kpc}$ for $D_\mathrm{star}$, following \citet{2017MNRAS.472..808R}.

Among the CC samples, RVs are available for 91 bright stars with $G\lesssim15~\mathrm{mag}$.
The mean RV is $150.25~\mathrm{km~s^{-1}}$, and the mean error of RV is $5.46~\mathrm{km~s^{-1}}$.
Here, RVs are referenced to the solar system barycentric frame.

\figref{CEP_CMD} shows ($G_\mathrm{BP}-G_\mathrm{RP}$, $G$) color-magnitude diagram (CMD) of the CC samples\footnote{Six CCs lack available $G_\mathrm{BP}-G_\mathrm{RP}$ values and are therefore not plotted in \figref{CEP_CMD}.}, along with the massive star candidates identified by \citet{2025arXiv250212251N}.
The CC samples are concentrated in a specific region of the CMD corresponding to the Cepheid instability strip, supporting the accuracy of the cross-matching. More distant CCs tend to be distributed toward the redder side of the CMD, likely reflects the increasing amount of intervening foreground dust along the line-of-sight.

Additionally, 38 CCs overlap with the massive star candidates\footnote{
There are five other massive Cepheids without available $D_\mathrm{star}$ (see \figref{CEP_CMD}), due to the saturation of the $V$ magnitudes provided by OGLE-IV \citep{2015AcA....65....1U, 2015AcA....65..297S}.}. 
These massive Cepheids can be regarded as gas tracers, satisfying $M_\mathrm{ini}\geq8M_\odot$ and $Age\lesssim\mathrm{50.0~Myr}$ \citep{2025arXiv250212251N}.
Among them, RVs are available for 37 objects. The mean RV calculated from only the massive Cepheids is $153.63~\mathrm{km~s^{-1}}$, with a mean RV error of $3.75~\mathrm{km~s^{-1}}$.

\subsection{Calculation of Internal PMs}
\label{sec:inPM}
For the calculation of internal PMs, we adopt a method similar to that of \citet{2025arXiv250212251N}. Specifically, we compute the internal PMs by subtracting the systemic PM of the SMC from the PMs of the CC samples.
However, \citet{2025arXiv250212251N} suggested that the spatial distribution and kinematics of massive star candidates with $Age\lesssim\mathrm{50.0~Myr}$ differ from those of older populations. It is not necessarily evident whether CCs with $Age\sim\mathrm{200~Myr}$ \citep{2017MNRAS.472..808R} follow the same kinematics as older populations.
Therefore, we adopt the mean PM of the CC samples as the systemic PM of the SMC for the calculation of their internal PMs.
The internal PMs calculated using the mean PM of the massive star candidates from \citet{2025arXiv250212251N} as the systemic PM are presented in \appref{sec:PMOB}.

\begin{figure*}
 \centering
 \includegraphics[width=18cm,clip]{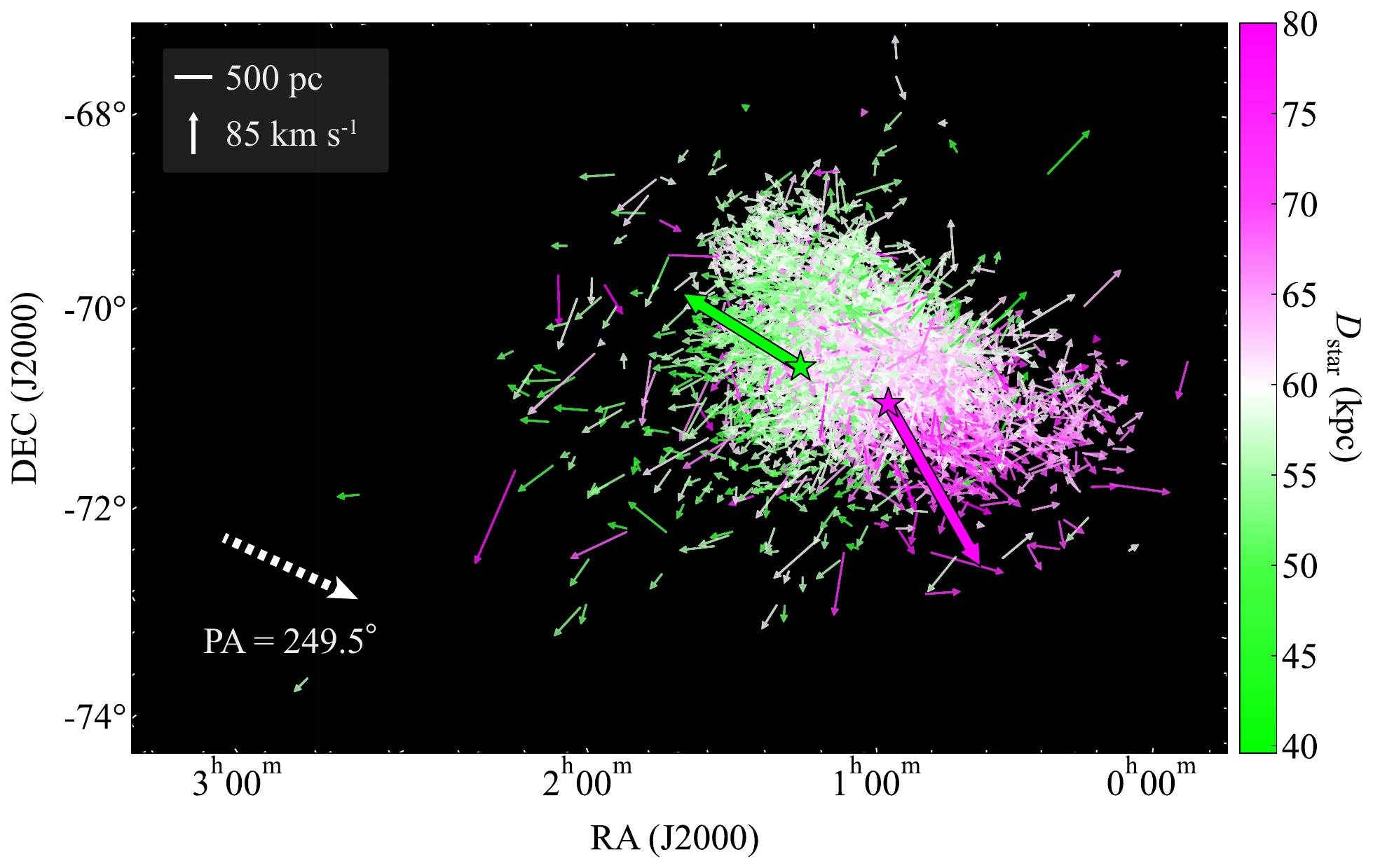}
 \caption{
  The internal PMs of the clipped CC samples determined by subtracting the mean PM of the clipped CC samples. 
  The legends indicate the magnitudes of PM vectors corresponding to $85~\mathrm{km~s^{-1}}$, representing the minimum escape velocity derived under the assumption of H\,\textsc{i} rotation \citep{2004ApJ...604..176S,2018NatAs...2..901M}.
  The colors of the PM vectors represent the distances $D_\mathrm{star}$ of the CCs, with more distant objects shown in magenta and nearer ones in green.
  The magenta and green star symbols indicate the mean positions of 241 distant ($>$70~$\mathrm{kpc}$) and 276 nearby ($<$55~$\mathrm{kpc}$) Cepheids, respectively.
  The thick vectors extending from the star symbols represent the mean PM vectors of the distant and nearby Cepheids, drawn at ten times the length of the individual stellar vectors for better visibility.
  Note that the distance thresholds for distant and nearby Cepheids ($70~\mathrm{kpc}$ and $55~\mathrm{kpc}$) are consistent with those used in \citet{2021MNRAS.502.2859N}.
  The white dotted arrow in the lower left of the figure indicates the direction in which the $D_\mathrm{star}$ gradient of the clipped CC samples is the largest.
}
\label{CEP_inPM}
\end{figure*}

To compute the mean PM, we select 4,105 objects from the CC samples that fall within $\pm2\sigma$ in both $\mu_{\alpha}$ and $\mu_{\delta}$. Hereafter, we refer to these 4,105 objects, to which the 2$\sigma$-cutoff is applied, as the clipped CC samples.
To account for coordinate distortions, we transform the coordinates to an orthographic projection following \citet{2018A&A...616A..12G}. For this transformation, we adopt the coordinate center as the center of the CC distribution, $(\alpha_\mathrm{cep},\,\delta_\mathrm{cep}) = (12.54^{\circ},\,-73.11^{\circ})$, derived by \citet{2017MNRAS.472..808R}.
Under the orthographic projection, the mean PM of the clipped CC samples is $(\mu_\alpha,\,\mu_\delta)=(0.705,\,-1.265)~\mathrm{mas~yr^{-1}}$.

This value is close to that derived by \citet{2021MNRAS.502.2859N}, which was obtained by assuming the center of the CC distribution based on data including older populations from the VMC survey, $(0.734 \pm 0.008,\,-1.234 \pm 0.008)~\mathrm{mas~yr^{-1}}$, but it does not agree within the error range.
Since the overall characteristics of the internal PMs remain unchanged regardless of which value is used, we adopt the mean PM of the clipped CC samples as the systemic PM for the internal PMs calculation.

Since the $D_\mathrm{star}$ for CCs are well determined, we subtract the mean PM using equation (6) of \citet{2025arXiv250212251N}.
After subtracting the mean PM, we perform an inverse transformation from the orthographic projection to restore the coordinates to the equatorial frame.
Through this procedure, we obtain the internal PMs of the CC samples, which are provided in \secref{sec:catalog}.

Note that, the magnitude of the mean PM corresponds to a tangential velocity of $\sim$400~$\mathrm{km~s^{-1}}$. Stars with extreme distances near 50~kpc or 80~kpc deviate by approximately 25\% from $D_\mathrm{SMC}$, meaning that neglecting $D_\mathrm{star}$ in the internal PMs calculation can introduce errors of up to $\sim$100~$\mathrm{km~s^{-1}}$.
This highlights the importance of accounting for $D_\mathrm{star}$ when computing the internal PMs of the CC samples.

\begin{figure*}
 \centering
 \includegraphics[width=18cm,clip]{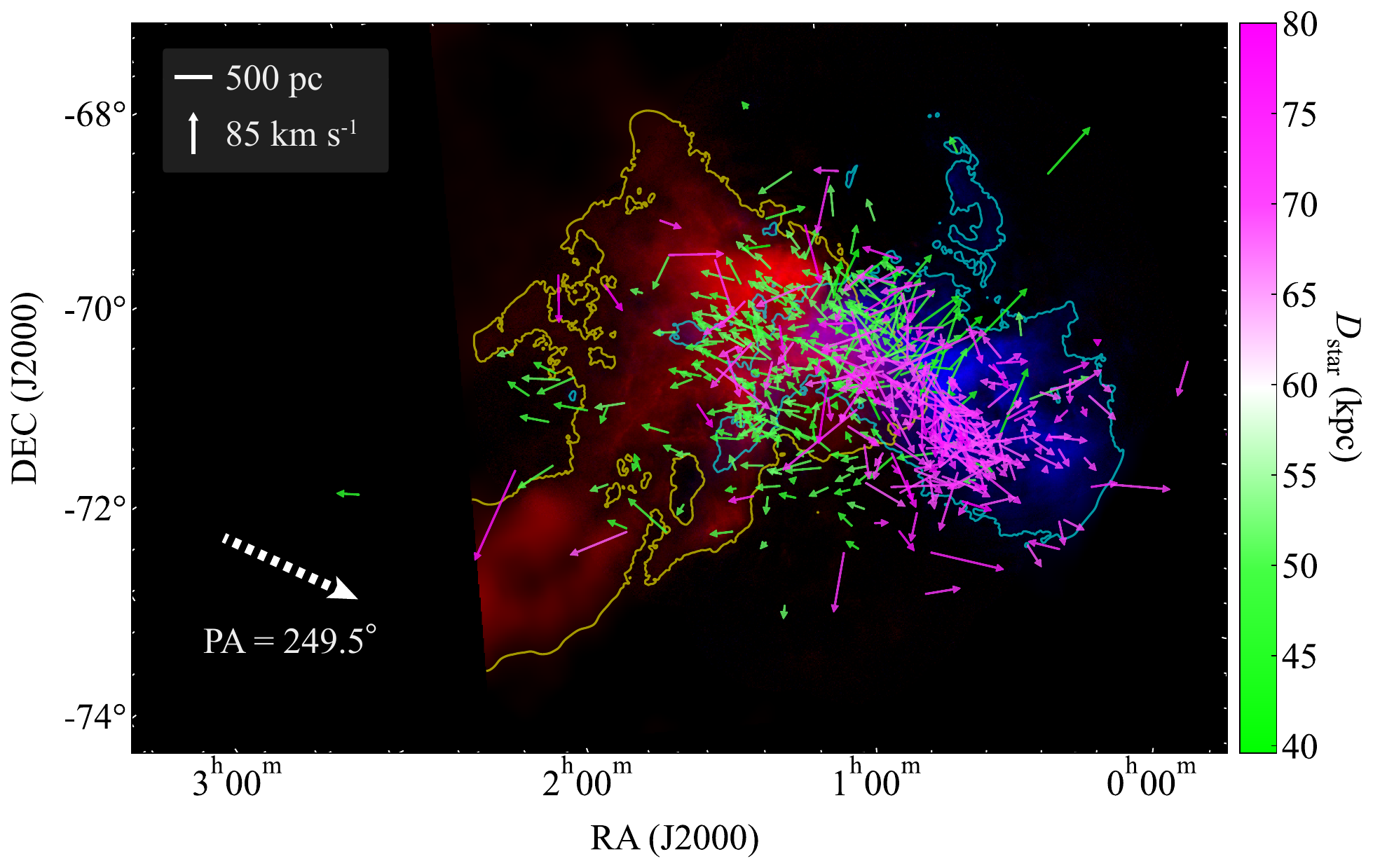}
 \caption{
 A comparison of the internal PMs of the clipped CC samples, considering only the 241 distant ($>$70~kpc) and 276 nearby ($<$55~kpc) populations, with the H\,\textsc{i} redshifted and blueshifted components.
 The magenta PM vectors represent the distant Cepheids ($>$70~kpc), while the green PM vectors represent the nearby Cepheids ($<$55~kpc).
 The background is a color multivelocity montage of the H\,\textsc{i} emission observed with ASKAP \& Parkes \citep{2018NatAs...2..901M}, where red represents the integrated intensity for the 179--187$~\mathrm{km~s^{-1}}$ range, and blue for the 105--113$~\mathrm{km~s^{-1}}$ range, both displayed on a common logarithmic scale of 10--1000$~\mathrm{K~km~s^{-1}}$.
 Contours highlight the regions where the H\,\textsc{i} redshifted and blueshifted components are present, with yellow contours corresponding to the redshifted component and cyan contours to the blueshifted component.
 Note that these velocity components are consistent with those shown in Figure~2 of \citet{2018NatAs...2..901M}.
 The white dotted arrow in the lower left of the figure indicates the direction in which the $D_\mathrm{star}$ gradient of the clipped CC samples is the largest.
}
\label{CEP_inPM_HI}
\end{figure*}

\begin{figure*}
 \centering
 \includegraphics[width=18cm,clip]{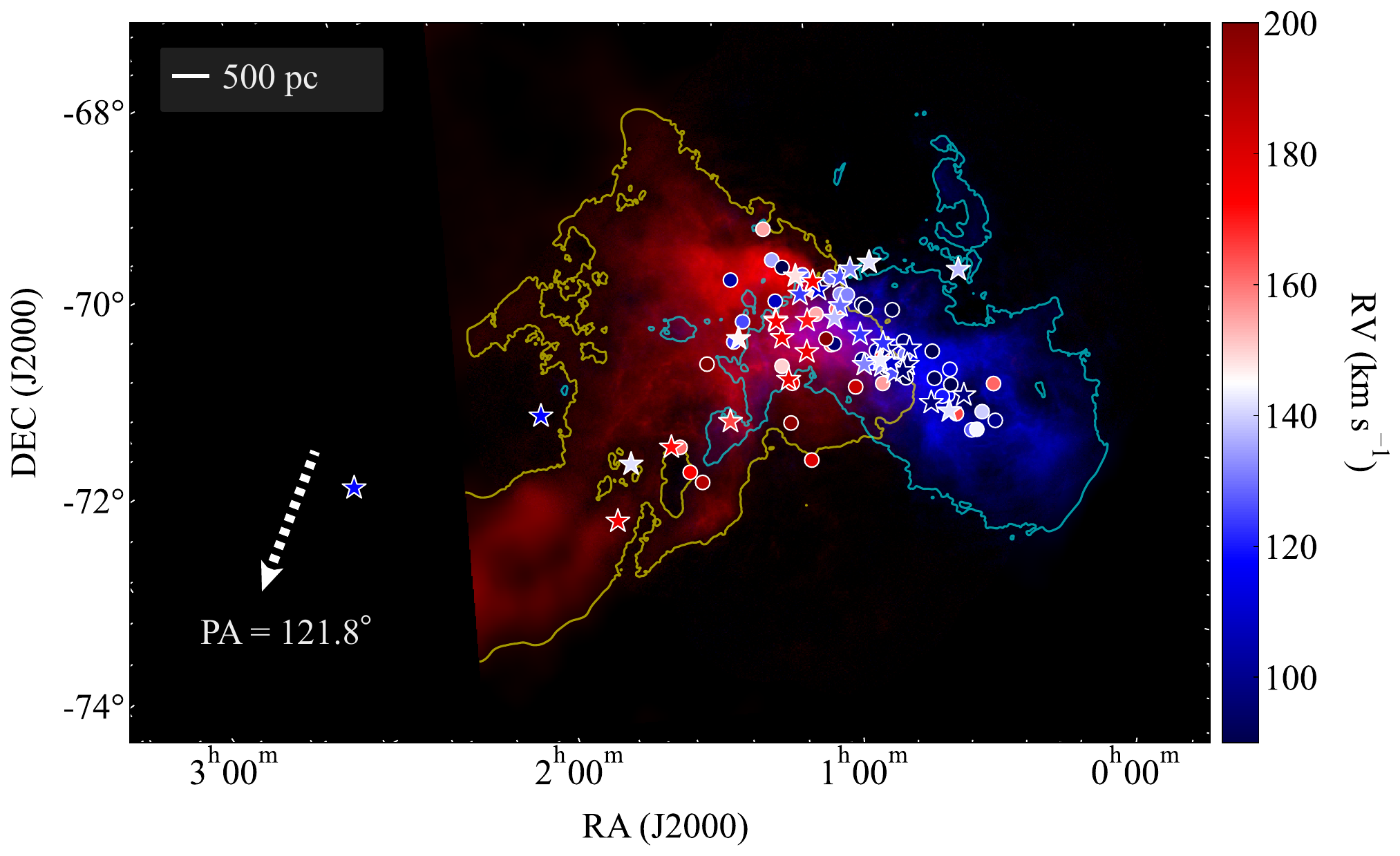}
 \caption{
 A comparison of the RVs of the 91 clipped CC samples with available RV data and the spatial distribution of the H\,\textsc{i} redshifted and blueshifted components.
 The star symbols represent 37 massive Cepheids that match the massive star candidates identified by \citet{2025arXiv250212251N} (see \figref{CEP_CMD}).
 The circles indicate the remaining 54 CCs. The colors of the star symbols and circles correspond to the RV values of the CCs.
 In this figure, the RVs are shown in reference to the LSRK, consistent with the line-of-sight velocities of the H\,\textsc{i}.
 The white dotted arrow in the lower left of the figure indicates the direction in which the RV gradient is the largest.
}
\label{CEP_RV}
\end{figure*}

\section{Results}
\label{sec:result}
\figref{CEP_inPM} shows the internal PMs and $D_\mathrm{star}$ of the clipped CC samples obtained in \secref{sec:sample}.  
As demonstrated by \citet{2017MNRAS.472..808R} and \citet{2021MNRAS.502.2859N}, the CCs exhibit a clear distance gradient extending from the northeast to the southwest \citep{1988ApJ...333..617M}.  
The position angle ($PA$) where the $D_\mathrm{star}$ gradient is most pronounced is $249.5^\circ$.  
Furthermore, our analysis reproduces the trend reported by \citet{2021MNRAS.502.2859N}, where nearer CCs move northeast relative to more distant ones.
Since our analysis focuses on internal PMs, it is evident that nearer CCs move northeast, while more distant CCs move southwest, almost in opposite directions.
In other words, in addition to the finding by \citet{2021MNRAS.502.2859N} that nearer CCs are moving away from the main body, we find that more distant CCs are also receding from it.
The outer CCs tend to move outward from the SMC, consistent with previous studies on various stellar populations based on internal PMs (e.g., \citealp{2018ApJ...864...55Z, 2018ApJ...867L...8O, 2019ApJ...887..267M, 2021MNRAS.502.2859N, 2025arXiv250212251N}).  
At the same time, the PMs of CCs do not exhibit a clear signature of galactic rotation.  
Only a few CCs are located in the southeastern Wing structure and the Magellanic Bridge.  
However, the small number of CCs distributed in the southeast move toward the LMC, suggesting an interaction with the LMC.

\figref{CEP_inPM_HI} shows the internal PMs of the clipped CC samples, specifically highlighting those located at distant ($>$70~kpc) and nearby ($<$55~kpc) regions.
The 241 distant CCs are predominantly found in the southwest, extending from the Bar, and generally move southwest. Those located in the northeast also exhibit motion toward the southwest.
In contrast, a small number of distant CCs near the southeastern Wing move toward the southeast. The westernmost distant CCs display somewhat random motions with relatively small magnitudes.
The 276 nearby CCs are widely distributed east of the SMC and tend to move eastward, with their northward velocity increasing toward higher declinations.
In the northeastern region, nearby CCs exhibit the prominent northeastward motion previously discovered by \citet{2021MNRAS.502.2859N}.
Both distant and nearby CCs include a distinct population located in the northwest that moves northwest.
When combined with the group moving southeast in the Wing, these findings indicate that the opposite motions in the eastern and western parts of the SMC, previously identified in massive star candidates by \citet{2025arXiv250212251N}, are also present in CCs.
This supports the interpretation, suggested by previous studies using internal PMs (e.g., \citealp{2018ApJ...864...55Z, 2018ApJ...867L...8O, 2019ApJ...887..267M, 2021MNRAS.502.2859N, 2025arXiv250212251N}), that the SMC is being destructively elongated by the tidal forces and/or ram pressure exerted by the LMC, located to the southeast.
The internal PM calculations accounting for the $D_\mathrm{star}$ of CCs demonstrate that the northwest-southeast elongation discovered by \citet{2025arXiv250212251N} is not merely a projection effect of distance. For further discussion on this elongation, including a comparison of the motions of massive star candidates and CCs, see also \appref{sec:PMOB}.

\figref{CEP_RV} shows the spatial distribution of the 91 clipped CC samples with available RVs, along with their RV values.
The distribution of stars with available RVs closely resembles that of the massive star candidates \citep{2025arXiv250212251N}, clearly outlining the Bar structure extending from the southwest to the north of the SMC and the Wing structure stretching from the north to the southeast.
This is explained by the fact that CCs with available RVs in \textit{Gaia} DR3 are bright stars with $G\lesssim15~\mathrm{mag}$, meaning that relatively young stars, which can serve as ISM tracers, have been selected.
The RVs of CCs decrease toward the northwest and increase toward the southeast.
The $PA$ at which the RV gradient is most pronounced is $121.8^\circ$.
This trend is consistent with the RV distributions of various stellar populations, including red giants and massive stars \citep{2008MNRAS.386..826E,2014MNRAS.442.1663D,2025arXiv250212251N}.

\begin{figure}
 \centering
 \includegraphics[width=8.5cm,clip]{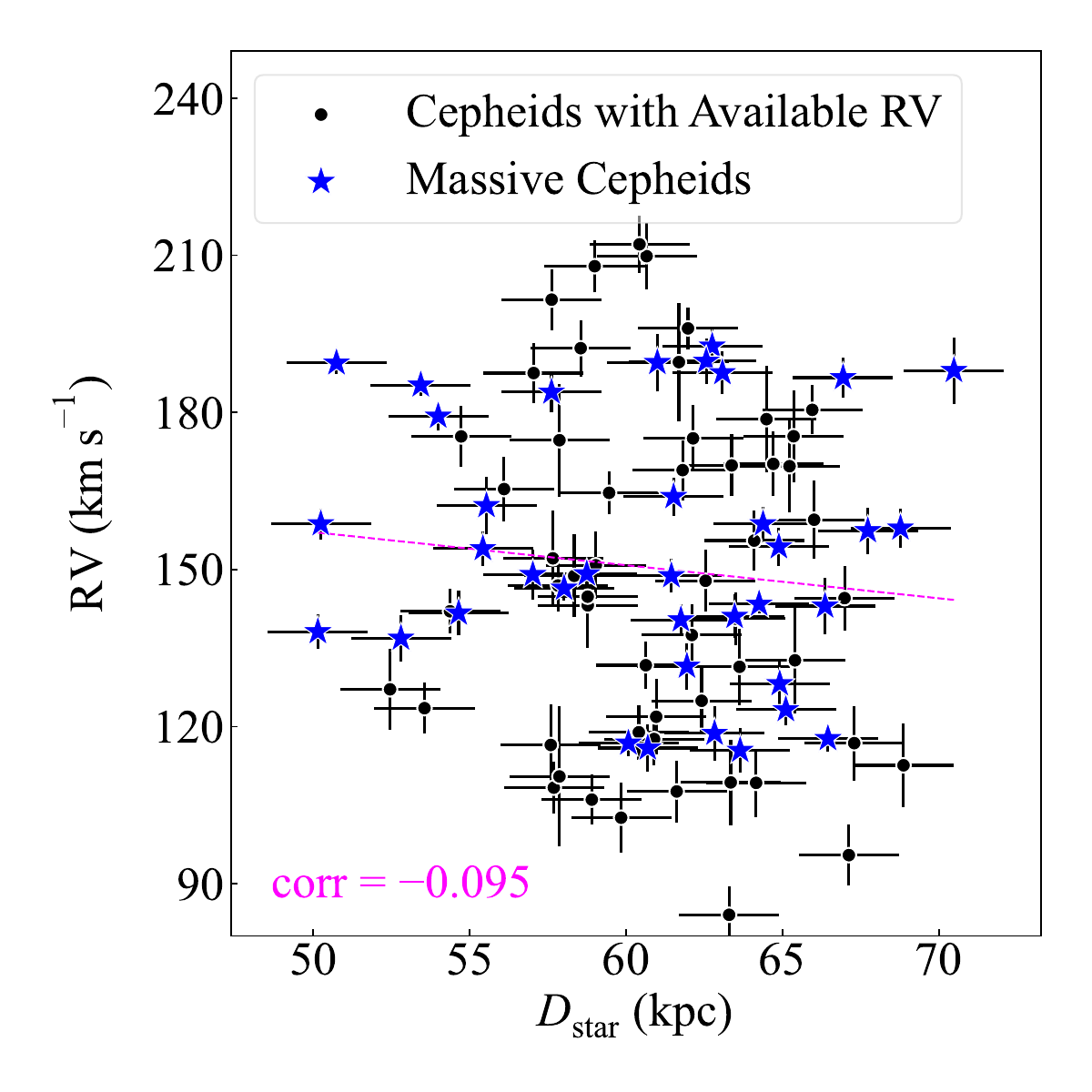}
 \caption{
 The relationship between $D_\mathrm{star}$ and RV for the 91 objects in the clipped CC samples with available RV measurements. The blue star symbols represent the 37 massive Cepheids that match the massive star candidates identified by \citet{2025arXiv250212251N} (see \figref{CEP_CMD}), while the black circles indicate the remaining 54 CCs. The magenta dashed line represents the result of a linear fit, with a correlation coefficient of $-0.095$ for the plotted data. The uncertainties in $D_\mathrm{star}$ are uniformly assumed to be 1.6~kpc \citep{2017MNRAS.472..808R}.
}
\label{CEP_D_RV}
\end{figure}

\section{Discussions}
\label{sec:discussion}

\subsection{Comparison of CC Distances and Neutral Hydrogen Gas Velocity Components}
\label{sec:HIcomp}
\figref{CEP_inPM_HI} compares the two velocity components of H\,\textsc{i} (a blueshifted component at approximately 100$~\mathrm{km~s^{-1}}$, and a redshifted component at approximately 200$~\mathrm{km~s^{-1}}$) with the internal PMs of the clipped CC samples, specifically focusing on the distant ($>$70~kpc) and nearby ($<$55~kpc) populations.
Distant CCs exhibit a spatial distribution similar to that of the H\,\textsc{i} blueshifted component, while nearby CCs resemble the H\,\textsc{i} redshifted component.
Furthermore, the velocity gradient of H\,\textsc{i} at $PA\sim60^\circ$ \citep{2004ApJ...604..176S, 2019MNRAS.483..392D} is nearly parallel to the $D_\mathrm{star}$ gradient of CCs at $PA=249.5^\circ$.
The agreement in spatial distribution and gradient suggests that these components are physically related structures; that is, the H\,\textsc{i} blueshifted component may be located farther away and moving southwest, while the H\,\textsc{i} redshifted component may be positioned nearer and moving northeast.
If this scenario is correct, the velocities of H\,\textsc{i} and the RVs of CCs should be consistent, meaning that nearer CCs should have larger RVs, while more distant CCs should have smaller RVs.
In the next section, we discuss the relationship between the $D_\mathrm{star}$ and RVs of CCs.

\subsection{No Correlation Between CCs' Distances and RVs}
\label{sec:D_RV}
\figref{CEP_D_RV} shows a plot of RV as a function of $D_\mathrm{star}$ for the 91 objects in the clipped CC sample with available RV measurements.
The correlation coefficient between $D_\mathrm{star}$ and RV is $-0.095$, indicating they are nearly uncorrelated.
This result remains unchanged even when considering only the 37 massive Cepheids, which serve as gas tracers\footnote{The correlation coefficient is $-0.095$ when calculated for only massive Cepheids, and $-0.097$ when calculated for the remaining CCs.}.
Thus, while the blueshifted/redshifted velocity components of H\,\textsc{i} and distant/nearby CCs appear to spatially coincide, they are kinematically inconsistent.
The lack of correlation between $D_\mathrm{star}$ and RV suggests that the two observed gradients arise from distinct physical processes.

The difference in $PA$ between the gradients of $D_\mathrm{star}$ and RVs supports the assertion that they originate from distinct mechanisms.
The $D_\mathrm{star}$ gradient of CCs is oriented in the northeast-southwest direction, with $PA = 249.5^{\circ}$.
In contrast, the RV gradient of CCs is oriented in the northwest-southeast direction, with $PA = 121.8^{\circ}$, indicating a clear discrepancy in their directions.
Since this RV gradient direction is consistent with previous studies \citep{2008MNRAS.386..826E,2014MNRAS.442.1663D}, it cannot be explained as an apparent gradient caused by the Wing structure detached from the main body of the SMC.

\citet{2012ApJ...750...36D} demonstrated that tidal forces from the LMC induce a northwest-southeast line-of-sight velocity gradient in the spheroidal component of the SMC, suggesting that the northwest-southeast gradient observed in the RVs of CCs can be interpreted as an effect of the interaction with the LMC.
Consequently, the northeast-southwest gradient seen in $D_\mathrm{star}$ of CCs may be a remnant of an earlier close encounter with the LMC. Alternatively, considering the $\sim$20--30~kpc line-of-sight depth of the SMC, which is aligned with the Milky Way, it may be influenced by interaction between the Milky Way and the SMC.

Various studies measuring the RVs of different stellar populations (e.g., \citealp{2008MNRAS.386..826E,2014MNRAS.442.1663D, 2025arXiv250212251N}) have pointed out that the velocity gradient of stars ($\sim$120$^\circ$) does not align with that of H\,\textsc{i} ($\sim$60$^\circ$), contradicting expectations from galactic rotation.
This discrepancy between the RV and H\,\textsc{i} velocity gradients closely resembles the inconsistency between the RV and  $D_\mathrm{star}$ gradients of CCs.
One possible explanation for this similarity is that while $D_\mathrm{star}$ of CCs traces the depth structure of H\,\textsc{i}, physical processes affecting only the gas—such as ram pressure or feedback from massive stars—have disrupted the velocity field of H\,\textsc{i}, leading to the observed mismatch between the RVs of stars and the velocity field of H\,\textsc{i}.
However, the fact that even the RVs of 37 massive Cepheids with $Age\lesssim\mathrm{50.0~Myr}$, as well as a larger number of massive star candidates identified by \citet{2025arXiv250212251N}, do not align with the H\,\textsc{i} velocity field poses a significant challenge to this explanation.
As previous studies on the internal PMs of stars in the SMC have suggested the SMC may lack galactic rotation (e.g., \citealp{2019ApJ...887..267M, 2021MNRAS.502.2859N, 2025arXiv250212251N}), this challenge could potentially be addressed through numerical simulations assuming a non-rotating SMC.

Furthermore, \figref{CEP_D_RV} does not show an increase in RVs proportional to $D_\mathrm{star}$, which \citet{2012ApJ...750...36D} cited as evidence for the Counter-Bridge (compare Figure~9 of \citet{2012ApJ...750...36D} with \figref{CEP_D_RV}).
Since the same result is obtained when considering only massive Cepheids, which serve as gas tracers, our findings appear robust.
This may be because the Counter-Bridge is a structure formed from a rotating disk.
If the SMC lacks galactic rotation, structures like the Counter-Bridge may not form.
Additionally, the Counter-Bridge is an arched structure extending toward the LMC, which further highlights its inconsistency with the linearly extended line-of-sight depth observed in CCs \citep{2017MNRAS.472..808R}.
The absence of evidence for the Counter-Bridge also underscores the need for new numerical simulations.
However, since the RV increase proportional to $D_\mathrm{star}$ in the Counter-Bridge becomes dominant at $\gtrsim$70~kpc, the lack of RV data for distant Cepheids in our sample may limit our ability to constrain the existence of the Counter-Bridge.
Moreover, in various stellar populations, stars with low-velocity ($\lesssim$150~$\mathrm{km~s^{-1}}$) RVs, extending to $\mathrm{RA} \approx 2\mathrm{h}$, exhibit an inverted V-shaped distribution, indicating that the RV distribution cannot be explained by a simple unidirectional velocity gradient.
This suggests the presence of a population with kinematics distinct from the main body of the SMC, which may be contaminating \figref{CEP_D_RV}.
For a more precise discussion, additional RV measurements from \textit{Gaia} DR4 are awaited.

We note that the massive Cepheids in \figref{CEP_D_RV} exhibit a broader $D_\mathrm{star}$ distribution compared to other CCs with available RVs, while their RV range appears narrower.
It is undeniable that young massive stars with $Age \lesssim \mathrm{50.0~Myr}$ also have a line-of-sight depth of $\sim$20~kpc.
The significant line-of-sight depth of the massive stars, which serve as gas tracers, suggests that H\,\textsc{i} likely possesses a similarly  line-of-sight depth.
To clarify the distances to the massive stars and the associated gas beyond the CCs, we are attempting to constrain the distances through isochrone fitting of star clusters containing a large number of massive stars, with the results to be reported in a forthcoming paper.

\section{Summary and Conclusions}
\label{sec:conclusion}
Using \textit{Gaia} DR3, we presented the distribution of internal PMs for 4,246 CCs, supported by their accurate distances, along with the relationship between $D_\mathrm{star}$ and RVs. The high-precision data from \textit{Gaia} DR3 increased the number of CCs with available RVs from 8 to 91 compared to \textit{Gaia} DR2. As a result, the internal PMs of CCs reproduced the motion reported by \citet{2021MNRAS.502.2859N}, where nearby CCs in the northeast move northeast relative to the distant CCs in the southwest. Furthermore, our internal PMs revealed that even the distant CCs move southwest relative to the main body. The distance gradient of the CCs is parallel to the line-of-sight velocity gradient of H\,\textsc{i}, but the RV gradient of the CCs is oriented differently from that of H\,\textsc{i}. The lack of correlation between the RVs and $D_\mathrm{star}$ of the CCs suggests that the two directional expansions indicated by their gradients were driven by different physical processes. Additionally, the CCs do not show the distance-dependent increase in RVs, which is observational evidence for the Counter-Bridge. Numerical simulations capable of reproducing these newly identified observational features are eagerly awaited.

\begin{acknowledgments}
This work was financially supported by JST SPRING, Grant Number
JPMJSP2125. The author SN would like to take this opportunity to thank the
“THERS Make New Standards Program for the Next Generation Researchers.”
KT is supported in part by JSPS KAKENHI Grant Number 20H01945.
SN would like to acknowledge the language assistance provided by ChatGPT (\url{https://chatgpt.com/}) and DeepL (\url{https://www.deepl.com/}) during the preparation of this manuscript. ChatGPT was particularly helpful in refining English expressions and enhancing the clarity of the text, while DeepL contributed to accurate and fluent translations. SN also wishes to express gratitude to OpenAI and DeepL SE for their technological contributions.
This work has made use of data from the European Space Agency (ESA) mission
{\it Gaia} (\url{https://www.cosmos.esa.int/gaia}), processed by the {\it Gaia}
Data Processing and Analysis Consortium (DPAC,
\url{https://www.cosmos.esa.int/web/gaia/dpac/consortium}). Funding for the DPAC
has been provided by national institutions, in particular the institutions
participating in the {\it Gaia} Multilateral Agreement.
This research has made use of the VizieR catalogue access tool, CDS,
Strasbourg, France \citep{10.26093/cds/vizier}. The original description 
of the VizieR service was published in \citet{vizier2000}.
This scientific work uses data obtained from Inyarrimanha Ilgari Bundara, the CSIRO Murchison Radio-astronomy Observatory. We acknowledge the Wajarri Yamaji People as the Traditional Owners and native title holders of the Observatory site. CSIRO’s ASKAP radio telescope and Murriyang, CSIRO's Parkes radio telescope, are part of the Australia Telescope National Facility (\url{https://ror.org/05qajvd42}). Operation of ASKAP is funded by the Australian Government with support from the National Collaborative Research Infrastructure Strategy. Parkes is funded by the Australian Government for operation as a National Facility managed by CSIRO. ASKAP uses the resources of the Pawsey Supercomputing Research Centre. Establishment of ASKAP, Inyarrimanha Ilgari Bundara, the CSIRO Murchison Radio-astronomy Observatory and the Pawsey Supercomputing Research Centre are initiatives of the Australian Government, with support from the Government of Western Australia and the Science and Industry Endowment Fund.
\end{acknowledgments}

%

\vspace{5mm}
\facilities{\textit{Gaia}, ASKAP, Parkes}


\software{aplpy \citep{2012ascl.soft08017R}, astropy \citep{2013A&A...558A..33A,2018AJ....156..123A}, CASA \citep{2007ASPC..376..127M}, ChatGPT \citep{2024arXiv241021276O}, dustapprox \citep{Fouesneau_dustapprox_2022}, matplotlib \citep{2007CSE.....9...90H}
}

\newpage

\appendix
\section{Catalog of 4,236 classical Cepheids in the SMC}
\label{sec:catalog}

\renewcommand{\arraystretch}{1.2}
\begin{deluxetable*}{lccccccccccc}[h]
 \tablecaption{Cross-match Results between \textit{Gaia} DR3 and OGLE-IV}
 \tablewidth{0pt}
 \label{catalog}
\tablehead{
    \shortstack{} &
    \shortstack{\textit{Gaia} DR3 ID} &
    \shortstack{OGLE ID} &
    \shortstack{RA (J2000)} &
    \shortstack{DEC (J2000)} &
    \shortstack{$G$} &
    \shortstack{$\sigma_{G}$} &
    \shortstack{...} &
    \shortstack{$D_\mathrm{star}$} &
    \shortstack{...} &\\
  &
  &
  &
 \colhead{(deg)}&
 \colhead{(deg)}&
 \colhead{(mag)}&
 \colhead{(mag)}&
 \colhead{ }&
 \colhead{(kpc)}&
 \colhead{ }&
}
\startdata
0 & 4688994540756312448 & OGLE-SMC-CEP-2476 & 13.99184887489 & -72.44262025649 & 19.706955 & 0.006493 & ... & 76.7 & ...\\
1 & 4689077656954612224 & OGLE-SMC-CEP-2095 & 13.42241613345 & -72.01297635880 & 18.745564 & 0.011079 & ... & 55.2 & ...\\
2 & 4688685577930565760 & OGLE-SMC-CEP-0022 & 5.89227970642 & -73.39970785978 & 19.047304 & 0.006729 & ... & 66.9 & ...\\
3 & 4687626365985065088 & OGLE-SMC-CEP-4548 & 19.41465531930 & -71.63650902714 & 18.945671 & 0.007297 & ... & 73.0 & ...\\
\enddata
\tablenotetext{ }{Here we only show part of the table. The machine readable table will be available in the HTML version of the article.}
\tablecomments{The complete table includes the following: the \textsc{source\_id} of \textit{Gaia} DR3 (\texttt{Gaia-DR3-ID}) and OGLE (\texttt{OGLE-ID}), coordinates in J2000 (\texttt{RAdeg}, \texttt{DEdeg}), photometry for $G$, $G_\mathrm{BP}$, and $G_\mathrm{RP}$ (\texttt{Gmag}, \texttt{GBPmag}, and \texttt{GRPmag}), the parallax and its uncertainty (\texttt{plx}), the \textit{Gaia} PMs (\texttt{pmRA}, \texttt{pmDE}), the distances $D_\mathrm{star}$ calculated assuming $D_\mathrm{SMC}=62~\mathrm{kpc}$ (\texttt{Dstar}; see \secref{sec:distance}), the \textit{Gaia} radial velocities and their uncertainties (\texttt{RVel}),  flags of the PM 2$\sigma$-cutoff samples (\texttt{PM-2s-clip}; see \secref{sec:inPM}), the internal PMs within the SMC calculated in \secref{sec:inPM} (\texttt{pmRA-in}, \texttt{pmDE-in}), the internal PMs respect to massive star candidates of \citet{2025arXiv250212251N} calculated in \appref{sec:PMOB} (\texttt{pmRA-in-OB}, \texttt{pmDE-in-OB}).}
\end{deluxetable*}
\renewcommand{\arraystretch}{1.0}

\begin{figure*}
 \centering
 \includegraphics[width=18cm,clip]{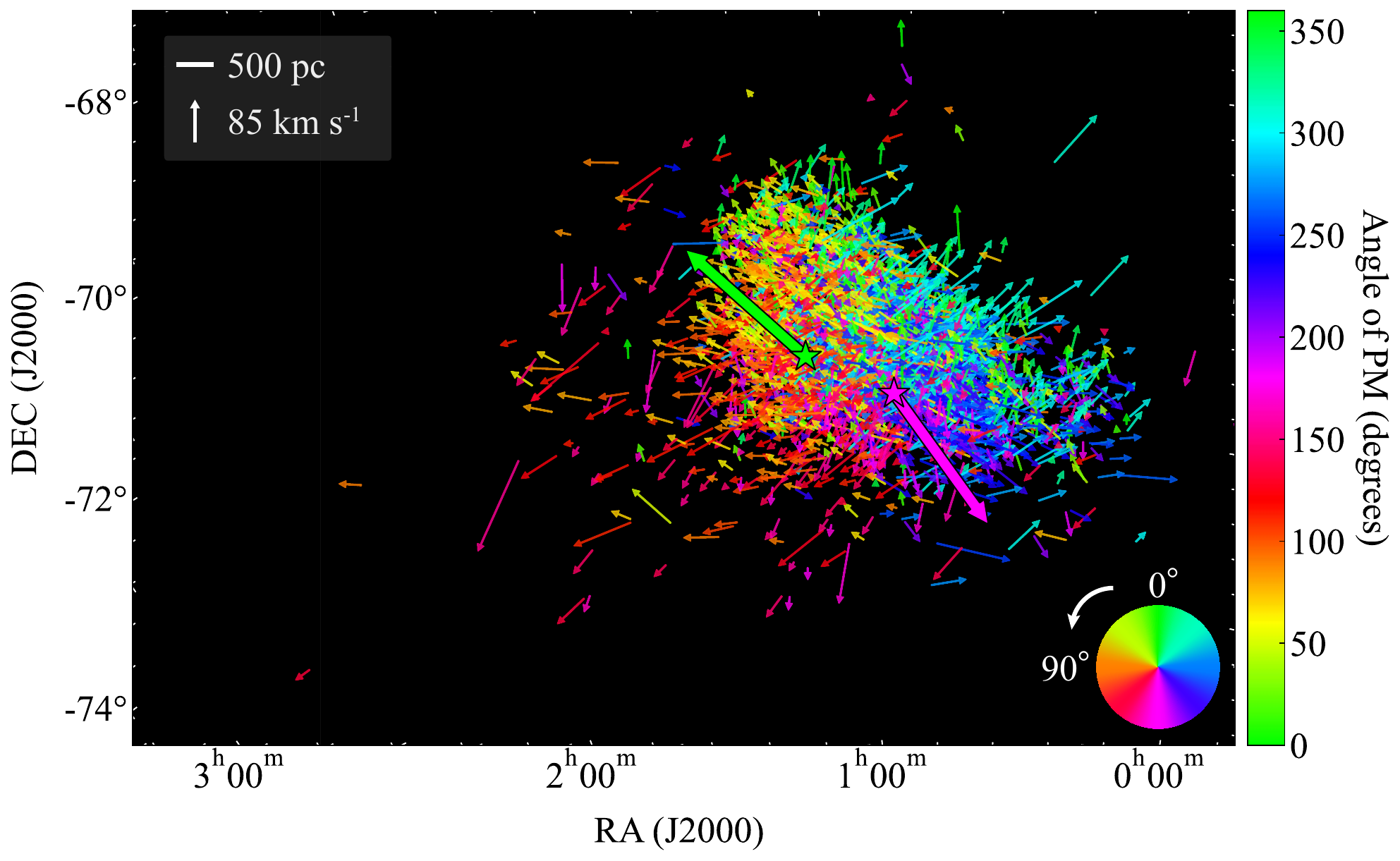}
 \caption{
 The internal PMs of the clipped CC samples, obtained by subtracting the mean PM of the massive star candidates \citep{2025arXiv250212251N}. The color of the PM vectors represents their direction, with the angle defined counterclockwise from north (0$^\circ$) in the ($x, y$) coordinates based on the the orthographic projection centered on the H\,\textsc{i} dynamical center \citep{2004ApJ...604..176S}. The magenta and green star symbols represent the mean positions of the distant ($>$70~kpc) and nearby ($<$55~kpc) Cepheids, respectively. The thick vectors extending from the star symbols indicate the mean PM vectors of the distant and nearby Cepheids, drawn at ten times the length of individual stellar vectors for better visibility.
}
\label{CEP_inPM_OB}
\end{figure*}

\begin{figure*}
 \centering
 \includegraphics[width=12cm,clip]{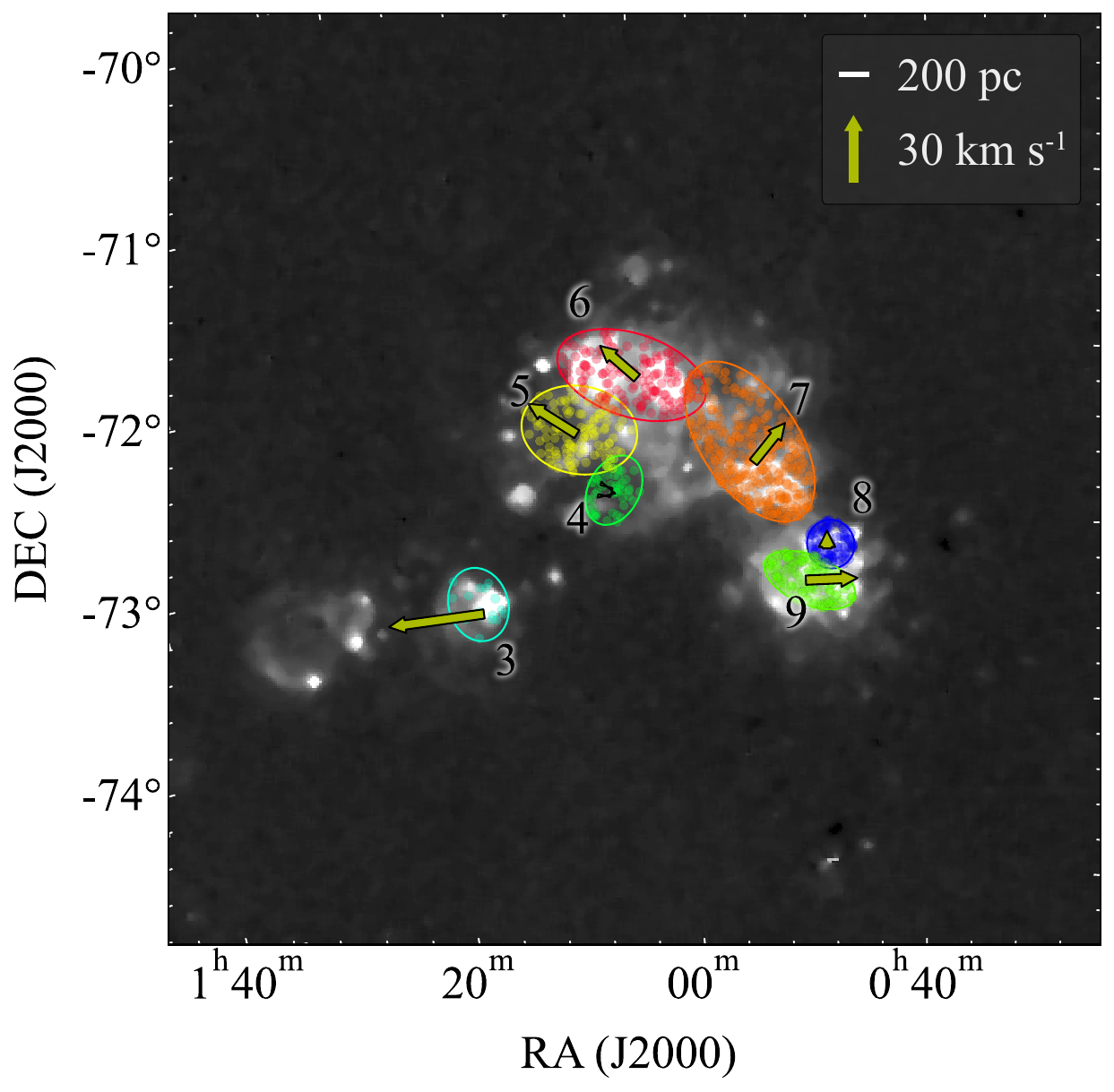}
 \caption{
 The median PMs of the clipped CC samples at locations corresponding to each Superstructure \citep{2025arXiv250212251N}. The median PMs are derived from the clipped CC samples located within the 2$\sigma$ elliptical regions, defined using the half-number radius of each Superstructure, as listed in Table~2 of \citet{2025arXiv250212251N}. Since the ellipses of Superstructure~1 and 2 contain only 0 and 3 CCs, respectively, their median values are not statistically significant and are therefore omitted from this figure. The background displays H$\alpha$ emission from SHASSA in the range of 0--600~deciRayleighs (dR; \citealp{2001PASP..113.1326G}).
}
\label{CEP_st9}
\end{figure*}

\section{Internal PMs of CCs with Respect to Young Massive Stars}
\label{sec:PMOB}
In this section, we compare the internal PMs of the massive star candidates from \citet{2025arXiv250212251N} with those of the CCs. In the calculations in \secref{sec:inPM}, we instead use the mean PM of the massive star candidates $(\mu_\alpha,\,\mu_\delta) = (0.785,\,-1.245)~\mathrm{mas~yr^{-1}}$ \citep{2025arXiv250212251N} for the systemic PM of the SMC, and the dynamical center of the H\,\textsc{i} gas $(\alpha_\mathrm{H_{I}},\,\delta_\mathrm{H_{I}})=(16.26^{\circ},\,-72.42^{\circ})$ \citep{2004ApJ...604..176S} as the coordinate center. A major difference is that the mean PM of the massive star candidates accounts for the motion of stars distributed in the Wing. The newly computed internal PMs of CCs referenced to the massive star candidates are listed in \secref{sec:catalog}.

\figref{CEP_inPM_OB} shows the internal PMs of the clipped CC samples obtained relative to the mean PM of the massive star candidates. The mean PMs of the distant and nearby CCs shown in \figref{CEP_inPM_OB} are more parallel compared to those shown in \figref{CEP_inPM}, making it clearer that they move in opposite directions along the distance gradient. This suggests that the near-far extension observed in the CCs \citep{2021MNRAS.502.2859N} occurs relative to the dynamical center of the H\,\textsc{i} gas, rather than the distribution center of the CCs \citep{2017MNRAS.472..808R}.

\figref{CEP_inPM_OB} is similar to the distribution of the internal PMs of massive star candidates shown in Figure~10 of \citet{2025arXiv250212251N}, with stars in the southeast moving southeast and stars in the northwest moving northwest. On the other hand, the northeast-southwest extension shown by \citet{2021MNRAS.502.2859N} is not observed in the massive star candidates and is a feature found only in the CCs.

\figref{CEP_st9} shows the median PMs of the CCs, which overlap with the positions of the Superstructures identified by \citet{2025arXiv250212251N}, which correspond to the $\sim$100~pc scale concentrations of massive stars classified by their surface density.
This provides an overview of the average internal PMs of the more central regions of the SMC. Similar to the motion of the Superstructures shown in Figure~14 of \citet{2025arXiv250212251N}, in Superstructure~3, a motion towards the LMC is observed, while in Superstructures~7 and 9, motions away from the LMC are seen.
The northwest-directed motion of Superstructure~7 highlights the dominant northwestward movement of the CCs in the northwest, regardless of distance, as shown in \figref{CEP_inPM_HI}. The lack of clear motion in Superstructure~8 and the westward motion in Superstructure~9 can be attributed to the dominant southwestward motion observed in the distant CCs, which are more concentrated in the southwest.

The CCs are located at approximately the same distance at the positions of Superstructure~5 and 6 (with medians of 58.0~kpc and 60.0~kpc, respectively), with both groups moving towards the northeast.
This contrasts with that reported by \citet{2025arXiv250212251N}, where different directions of motion were observed for Superstructure~4, 5, and 6 under the assumption that all the massive star candidates are at the same distance.
This suggests that either young massive star candidates in Superstructures~5 and 6 are at different distances, or they are at the same distance but exhibit different motions. In either case, at the positions of Superstructures~5 and 6, the CCs and the massive star candidates exhibit different properties. This fact supports assertion of \citet{2025arXiv250212251N} that the distribution and motion of massive star candidates differ from those of the older population.

Note that there are more massive star candidates with larger PMs that would escape from the SMC. This suggests the possibility that recent star formation is occurring in gas that has escaped the SMC due to the tidal interaction, ram pressure or feedback. Furthermore, while the CCs are distributed in the southern region of the SMC, neither the massive star candidates nor the ISM are found in the southern region of the SMC. This separation may imply that not only tidal forces from the LMC but also ram pressure and/or feedback have acted towards the LMC direction. In \figref{CEP_inPM_OB}, the internal PMs of the CCs located in the southern SMC are polarized into two directions—southwest and southeast—indicating that stars being stripped from the bar are moving southeast.


\end{CJK}

\begin{thebibliography}{}
\expandafter\ifx\csname natexlab\endcsname\relax\def\natexlab#1{#1}\fi
\providecommand{\url}[1]{\href{#1}{#1}}
\providecommand{\dodoi}[1]{doi:~\href{http://doi.org/#1}{\nolinkurl{#1}}}
\providecommand{\doeprint}[1]{\href{http://ascl.net/#1}{\nolinkurl{http://ascl.net/#1}}}
\providecommand{\doarXiv}[1]{\href{https://arxiv.org/abs/#1}{\nolinkurl{https://arxiv.org/abs/#1}}}

\bibitem[{ {Astropy Collaboration} {et~al.}(2013){Astropy Collaboration}, {Robitaille}, {Tollerud}, {Greenfield}, {Droettboom}, {Bray}, {Aldcroft}, {Davis}, {Ginsburg}, {Price-Whelan}, {Kerzendorf}, {Conley}, {Crighton}, {Barbary}, {Muna}, {Ferguson}, {Grollier}, {Parikh}, {Nair}, {Unther}, {Deil}, {Woillez}, {Conseil}, {Kramer}, {Turner}, {Singer}, {Fox}, {Weaver}, {Zabalza}, {Edwards}, {Azalee Bostroem}, {Burke}, {Casey}, {Crawford}, {Dencheva}, {Ely}, {Jenness}, {Labrie}, {Lim}, {Pierfederici}, {Pontzen}, {Ptak}, {Refsdal}, {Servillat}, \& {Streicher}}]{2013A&A...558A..33A}
{Astropy Collaboration}, {Robitaille}, T.~P., {Tollerud}, E.~J., {et~al.} 2013, \bibinfo{title}{{Astropy: A community Python package for astronomy},} \aap, 558, A33, \dodoi{10.1051/0004-6361/201322068}

\bibitem[{ {Astropy Collaboration} {et~al.}(2018){Astropy Collaboration}, {Price-Whelan}, {Sip{\H{o}}cz}, {G{\"u}nther}, {Lim}, {Crawford}, {Conseil}, {Shupe}, {Craig}, {Dencheva}, {Ginsburg}, {VanderPlas}, {Bradley}, {P{\'e}rez-Su{\'a}rez}, {de Val-Borro}, {Aldcroft}, {Cruz}, {Robitaille}, {Tollerud}, {Ardelean}, {Babej}, {Bach}, {Bachetti}, {Bakanov}, {Bamford}, {Barentsen}, {Barmby}, {Baumbach}, {Berry}, {Biscani}, {Boquien}, {Bostroem}, {Bouma}, {Brammer}, {Bray}, {Breytenbach}, {Buddelmeijer}, {Burke}, {Calderone}, {Cano Rodr{\'\i}guez}, {Cara}, {Cardoso}, {Cheedella}, {Copin}, {Corrales}, {Crichton}, {D'Avella}, {Deil}, {Depagne}, {Dietrich}, {Donath}, {Droettboom}, {Earl}, {Erben}, {Fabbro}, {Ferreira}, {Finethy}, {Fox}, {Garrison}, {Gibbons}, {Goldstein}, {Gommers}, {Greco}, {Greenfield}, {Groener}, {Grollier}, {Hagen}, {Hirst}, {Homeier}, {Horton}, {Hosseinzadeh}, {Hu}, {Hunkeler}, {Ivezi{\'c}}, {Jain}, {Jenness}, {Kanarek}, {Kendrew}, {Kern}, {Kerzendorf}, {Khvalko}, {King}, {Kirkby}, {Kulkarni},
  {Kumar}, {Lee}, {Lenz}, {Littlefair}, {Ma}, {Macleod}, {Mastropietro}, {McCully}, {Montagnac}, {Morris}, {Mueller}, {Mumford}, {Muna}, {Murphy}, {Nelson}, {Nguyen}, {Ninan}, {N{\"o}the}, {Ogaz}, {Oh}, {Parejko}, {Parley}, {Pascual}, {Patil}, {Patil}, {Plunkett}, {Prochaska}, {Rastogi}, {Reddy Janga}, {Sabater}, {Sakurikar}, {Seifert}, {Sherbert}, {Sherwood-Taylor}, {Shih}, {Sick}, {Silbiger}, {Singanamalla}, {Singer}, {Sladen}, {Sooley}, {Sornarajah}, {Streicher}, {Teuben}, {Thomas}, {Tremblay}, {Turner}, {Terr{\'o}n}, {van Kerkwijk}, {de la Vega}, {Watkins}, {Weaver}, {Whitmore}, {Woillez}, {Zabalza}, \& {Astropy Contributors}}]{2018AJ....156..123A}
{Astropy Collaboration}, {Price-Whelan}, A.~M., {Sip{\H{o}}cz}, B.~M., {et~al.} 2018, \bibinfo{title}{{The Astropy Project: Building an Open-science Project and Status of the v2.0 Core Package},} \aj, 156, 123, \dodoi{10.3847/1538-3881/aabc4f}

\bibitem[{G. {Besla} {et~al.}(2007){Besla}, {Kallivayalil}, {Hernquist}, {Robertson}, {Cox}, {van der Marel}, \& {Alcock}}]{2007ApJ...668..949B}
{Besla}, G., {Kallivayalil}, N., {Hernquist}, L., {et~al.} 2007, \bibinfo{title}{{Are the Magellanic Clouds on Their First Passage about the Milky Way?},} \apj, 668, 949, \dodoi{10.1086/521385}

\bibitem[{G. {Besla} {et~al.}(2012){Besla}, {Kallivayalil}, {Hernquist}, {van der Marel}, {Cox}, \& {Kere{\v{s}}}}]{2012MNRAS.421.2109B}
{Besla}, G., {Kallivayalil}, N., {Hernquist}, L., {et~al.} 2012, \bibinfo{title}{{The role of dwarf galaxy interactions in shaping the Magellanic System and implications for Magellanic Irregulars},} \mnras, 421, 2109, \dodoi{10.1111/j.1365-2966.2012.20466.x}

\bibitem[{A. {Bressan} {et~al.}(2012){Bressan}, {Marigo}, {Girardi}, {Salasnich}, {Dal Cero}, {Rubele}, \& {Nanni}}]{2012MNRAS.427..127B}
{Bressan}, A., {Marigo}, P., {Girardi}, L., {et~al.} 2012, \bibinfo{title}{{PARSEC: stellar tracks and isochrones with the PAdova and TRieste Stellar Evolution Code},} \mnras, 427, 127, \dodoi{10.1111/j.1365-2966.2012.21948.x}

\bibitem[{Y. {Chen} {et~al.}(2015){Chen}, {Bressan}, {Girardi}, {Marigo}, {Kong}, \& {Lanza}}]{2015MNRAS.452.1068C}
{Chen}, Y., {Bressan}, A., {Girardi}, L., {et~al.} 2015, \bibinfo{title}{{PARSEC evolutionary tracks of massive stars up to 350 M$_{{\ensuremath{\odot}}}$ at metallicities 0.0001 {\ensuremath{\leq}} Z {\ensuremath{\leq}} 0.04},} \mnras, 452, 1068, \dodoi{10.1093/mnras/stv1281}

\bibitem[{Y. {Chen} {et~al.}(2014){Chen}, {Girardi}, {Bressan}, {Marigo}, {Barbieri}, \& {Kong}}]{2014MNRAS.444.2525C}
{Chen}, Y., {Girardi}, L., {Bressan}, A., {et~al.} 2014, \bibinfo{title}{{Improving PARSEC models for very low mass stars},} \mnras, 444, 2525, \dodoi{10.1093/mnras/stu1605}

\bibitem[{M.~R.~L. {Cioni} {et~al.}(2011){Cioni}, {Clementini}, {Girardi}, {Guandalini}, {Gullieuszik}, {Miszalski}, {Moretti}, {Ripepi}, {Rubele}, {Bagheri}, {Bekki}, {Cross}, {de Blok}, {de Grijs}, {Emerson}, {Evans}, {Gibson}, {Gonzales-Solares}, {Groenewegen}, {Irwin}, {Ivanov}, {Lewis}, {Marconi}, {Marquette}, {Mastropietro}, {Moore}, {Napiwotzki}, {Naylor}, {Oliveira}, {Read}, {Sutorius}, {van Loon}, {Wilkinson}, \& {Wood}}]{2011A&A...527A.116C}
{Cioni}, M. R.~L., {Clementini}, G., {Girardi}, L., {et~al.} 2011, \bibinfo{title}{{The VMC survey. I. Strategy and first data},} \aap, 527, A116, \dodoi{10.1051/0004-6361/201016137}

\bibitem[{R. {de Grijs} \& G. {Bono}(2015){de Grijs} \& {Bono}}]{2015AJ....149..179D}
{de Grijs}, R., \& {Bono}, G. 2015, \bibinfo{title}{{Clustering of Local Group Distances: Publication Bias or Correlated Measurements? III. The Small Magellanic Cloud},} \aj, 149, 179, \dodoi{10.1088/0004-6256/149/6/179}

\bibitem[{R. {de Grijs} {et~al.}(2014){de Grijs}, {Wicker}, \& {Bono}}]{2014AJ....147..122D}
{de Grijs}, R., {Wicker}, J.~E., \& {Bono}, G. 2014, \bibinfo{title}{{Clustering of Local Group Distances: Publication Bias or Correlated Measurements? I. The Large Magellanic Cloud},} \aj, 147, 122, \dodoi{10.1088/0004-6256/147/5/122}

\bibitem[{M. {De Leo} {et~al.}(2020){De Leo}, {Carrera}, {No{\"e}l}, {Read}, {Erkal}, \& {Gallart}}]{2020MNRAS.495...98D}
{De Leo}, M., {Carrera}, R., {No{\"e}l}, N. E.~D., {et~al.} 2020, \bibinfo{title}{{Revealing the tidal scars of the Small Magellanic Cloud},} \mnras, 495, 98, \dodoi{10.1093/mnras/staa1122}

\bibitem[{E.~M. {Di Teodoro} {et~al.}(2019){Di Teodoro}, {McClure-Griffiths}, {Jameson}, {D{\'e}nes}, {Dickey}, {Stanimirovi{\'c}}, {Staveley-Smith}, {Anderson}, {Bunton}, {Chippendale}, {Lee-Waddell}, {MacLeod}, \& {Voronkov}}]{2019MNRAS.483..392D}
{Di Teodoro}, E.~M., {McClure-Griffiths}, N.~M., {Jameson}, K.~E., {et~al.} 2019, \bibinfo{title}{{On the dynamics of the Small Magellanic Cloud through high-resolution ASKAP H I observations},} \mnras, 483, 392, \dodoi{10.1093/mnras/sty3095}

\bibitem[{J.~D. {Diaz} \& K. {Bekki}(2012){Diaz} \& {Bekki}}]{2012ApJ...750...36D}
{Diaz}, J.~D., \& {Bekki}, K. 2012, \bibinfo{title}{{The Tidal Origin of the Magellanic Stream and the Possibility of a Stellar Counterpart},} \apj, 750, 36, \dodoi{10.1088/0004-637X/750/1/36}

\bibitem[{P.~D. {Dobbie} {et~al.}(2014){Dobbie}, {Cole}, {Subramaniam}, \& {Keller}}]{2014MNRAS.442.1663D}
{Dobbie}, P.~D., {Cole}, A.~A., {Subramaniam}, A., \& {Keller}, S. 2014, \bibinfo{title}{{Red giants in the Small Magellanic Cloud - I. Disc and tidal stream kinematics},} \mnras, 442, 1663, \dodoi{10.1093/mnras/stu910}

\bibitem[{C.~J. {Evans} \& I.~D. {Howarth}(2008){Evans} \& {Howarth}}]{2008MNRAS.386..826E}
{Evans}, C.~J., \& {Howarth}, I.~D. 2008, \bibinfo{title}{{Kinematics of massive stars in the Small Magellanic Cloud},} \mnras, 386, 826, \dodoi{10.1111/j.1365-2966.2008.13012.x}

\bibitem[{M. Fouesneau {et~al.}(2022)Fouesneau, Andrae, Sordo, \& Dharmawardena}]{Fouesneau_dustapprox_2022}
Fouesneau, M., Andrae, R., Sordo, R., \& Dharmawardena, T. 2022, \bibinfo{title}{{dustapprox},}, 0.1 \url{https://github.com/mfouesneau/dustapprox}

\bibitem[{ {Gaia Collaboration} {et~al.}(2016){Gaia Collaboration}, {Prusti}, {de Bruijne}, {Brown}, {Vallenari}, {Babusiaux}, {Bailer-Jones}, {Bastian}, {Biermann}, {Evans}, {Eyer}, {Jansen}, {Jordi}, {Klioner}, {Lammers}, {Lindegren}, {Luri}, {Mignard}, {Milligan}, {Panem}, {Poinsignon}, {Pourbaix}, {Randich}, {Sarri}, {Sartoretti}, {Siddiqui}, {Soubiran}, {Valette}, {van Leeuwen}, {Walton}, {Aerts}, {Arenou}, {Cropper}, {Drimmel}, {H{\o}g}, {Katz}, {Lattanzi}, {O'Mullane}, {Grebel}, {Holland}, {Huc}, {Passot}, {Bramante}, {Cacciari}, {Casta{\~n}eda}, {Chaoul}, {Cheek}, {De Angeli}, {Fabricius}, {Guerra}, {Hern{\'a}ndez}, {Jean-Antoine-Piccolo}, {Masana}, {Messineo}, {Mowlavi}, {Nienartowicz}, {Ord{\'o}{\~n}ez-Blanco}, {Panuzzo}, {Portell}, {Richards}, {Riello}, {Seabroke}, {Tanga}, {Th{\'e}venin}, {Torra}, {Els}, {Gracia-Abril}, {Comoretto}, {Garcia-Reinaldos}, {Lock}, {Mercier}, {Altmann}, {Andrae}, {Astraatmadja}, {Bellas-Velidis}, {Benson}, {Berthier}, {Blomme}, {Busso}, {Carry}, {Cellino},
  {Clementini}, {Cowell}, {Creevey}, {Cuypers}, {Davidson}, {De Ridder}, {de Torres}, {Delchambre}, {Dell'Oro}, {Ducourant}, {Fr{\'e}mat}, {Garc{\'\i}a-Torres}, {Gosset}, {Halbwachs}, {Hambly}, {Harrison}, {Hauser}, {Hestroffer}, {Hodgkin}, {Huckle}, {Hutton}, {Jasniewicz}, {Jordan}, {Kontizas}, {Korn}, {Lanzafame}, {Manteiga}, {Moitinho}, {Muinonen}, {Osinde}, {Pancino}, {Pauwels}, {Petit}, {Recio-Blanco}, {Robin}, {Sarro}, {Siopis}, {Smith}, {Smith}, {Sozzetti}, {Thuillot}, {van Reeven}, {Viala}, {Abbas}, {Abreu Aramburu}, {Accart}, {Aguado}, {Allan}, {Allasia}, {Altavilla}, {{\'A}lvarez}, {Alves}, {Anderson}, {Andrei}, {Anglada Varela}, {Antiche}, {Antoja}, {Ant{\'o}n}, {Arcay}, {Atzei}, {Ayache}, {Bach}, {Baker}, {Balaguer-N{\'u}{\~n}ez}, {Barache}, {Barata}, {Barbier}, {Barblan}, {Baroni}, {Barrado y Navascu{\'e}s}, {Barros}, {Barstow}, {Becciani}, {Bellazzini}, {Bellei}, {Bello Garc{\'\i}a}, {Belokurov}, {Bendjoya}, {Berihuete}, {Bianchi}, {Bienaym{\'e}}, {Billebaud}, {Blagorodnova}, {Blanco-Cuaresma},
  {Boch}, {Bombrun}, {Borrachero}, {Bouquillon}, {Bourda}, {Bouy}, {Bragaglia}, {Breddels}, {Brouillet}, {Br{\"u}semeister}, {Bucciarelli}, {Budnik}, {Burgess}, {Burgon}, {Burlacu}, {Busonero}, {Buzzi}, {Caffau}, {Cambras}, {Campbell}, {Cancelliere}, {Cantat-Gaudin}, {Carlucci}, {Carrasco}, {Castellani}, {Charlot}, {Charnas}, {Charvet}, {Chassat}, {Chiavassa}, {Clotet}, {Cocozza}, {Collins}, {Collins}, {Costigan}, {Crifo}, {Cross}, {Crosta}, {Crowley}, {Dafonte}, {Damerdji}, {Dapergolas}, {David}, {David}, {De Cat}, {de Felice}, {de Laverny}, {De Luise}, {De March}, {de Martino}, {de Souza}, {Debosscher}, {del Pozo}, {Delbo}, {Delgado}, {Delgado}, {di Marco}, {Di Matteo}, {Diakite}, {Distefano}, {Dolding}, {Dos Anjos}, {Drazinos}, {Dur{\'a}n}, {Dzigan}, {Ecale}, {Edvardsson}, {Enke}, {Erdmann}, {Escolar}, {Espina}, {Evans}, {Eynard Bontemps}, {Fabre}, {Fabrizio}, {Faigler}, {Falc{\~a}o}, {Farr{\`a}s Casas}, {Faye}, {Federici}, {Fedorets}, {Fern{\'a}ndez-Hern{\'a}ndez}, {Fernique}, {Fienga}, {Figueras},
  {Filippi}, {Findeisen}, {Fonti}, {Fouesneau}, {Fraile}, {Fraser}, {Fuchs}, {Furnell}, {Gai}, {Galleti}, {Galluccio}, {Garabato}, {Garc{\'\i}a-Sedano}, {Gar{\'e}}, {Garofalo}, {Garralda}, {Gavras}, {Gerssen}, {Geyer}, {Gilmore}, {Girona}, {Giuffrida}, {Gomes}, {Gonz{\'a}lez-Marcos}, {Gonz{\'a}lez-N{\'u}{\~n}ez}, {Gonz{\'a}lez-Vidal}, {Granvik}, {Guerrier}, {Guillout}, {Guiraud}, {G{\'u}rpide}, {Guti{\'e}rrez-S{\'a}nchez}, {Guy}, {Haigron}, {Hatzidimitriou}, {Haywood}, {Heiter}, {Helmi}, {Hobbs}, {Hofmann}, {Holl}, {Holland}, {Hunt}, {Hypki}, {Icardi}, {Irwin}, {Jevardat de Fombelle}, {Jofr{\'e}}, {Jonker}, {Jorissen}, {Julbe}, {Karampelas}, {Kochoska}, {Kohley}, {Kolenberg}, {Kontizas}, {Koposov}, {Kordopatis}, {Koubsky}, {Kowalczyk}, {Krone-Martins}, {Kudryashova}, {Kull}, {Bachchan}, {Lacoste-Seris}, {Lanza}, {Lavigne}, {Le Poncin-Lafitte}, {Lebreton}, {Lebzelter}, {Leccia}, {Leclerc}, {Lecoeur-Taibi}, {Lemaitre}, {Lenhardt}, {Leroux}, {Liao}, {Licata}, {Lindstr{\o}m}, {Lister}, {Livanou}, {Lobel},
  {L{\"o}ffler}, {L{\'o}pez}, {Lopez-Lozano}, {Lorenz}, {Loureiro}, {MacDonald}, {Magalh{\~a}es Fernandes}, {Managau}, {Mann}, {Mantelet}, {Marchal}, {Marchant}, {Marconi}, {Marie}, {Marinoni}, {Marrese}, {Marschalk{\'o}}, {Marshall}, {Mart{\'\i}n-Fleitas}, {Martino}, {Mary}, {Matijevi{\v{c}}}, {Mazeh}, {McMillan}, {Messina}, {Mestre}, {Michalik}, {Millar}, {Miranda}, {Molina}, {Molinaro}, {Molinaro}, {Moln{\'a}r}, {Moniez}, {Montegriffo}, {Monteiro}, {Mor}, {Mora}, {Morbidelli}, {Morel}, {Morgenthaler}, {Morley}, {Morris}, {Mulone}, {Muraveva}, {Musella}, {Narbonne}, {Nelemans}, {Nicastro}, {Noval}, {Ord{\'e}novic}, {Ordieres-Mer{\'e}}, {Osborne}, {Pagani}, {Pagano}, {Pailler}, {Palacin}, {Palaversa}, {Parsons}, {Paulsen}, {Pecoraro}, {Pedrosa}, {Pentik{\"a}inen}, {Pereira}, {Pichon}, {Piersimoni}, {Pineau}, {Plachy}, {Plum}, {Poujoulet}, {Pr{\v{s}}a}, {Pulone}, {Ragaini}, {Rago}, {Rambaux}, {Ramos-Lerate}, {Ranalli}, {Rauw}, {Read}, {Regibo}, {Renk}, {Reyl{\'e}}, {Ribeiro}, {Rimoldini}, {Ripepi}, {Riva},
  {Rixon}, {Roelens}, {Romero-G{\'o}mez}, {Rowell}, {Royer}, {Rudolph}, {Ruiz-Dern}, {Sadowski}, {Sagrist{\`a} Sell{\'e}s}, {Sahlmann}, {Salgado}, {Salguero}, {Sarasso}, {Savietto}, {Schnorhk}, {Schultheis}, {Sciacca}, {Segol}, {Segovia}, {Segransan}, {Serpell}, {Shih}, {Smareglia}, {Smart}, {Smith}, {Solano}, {Solitro}, {Sordo}, {Soria Nieto}, {Souchay}, {Spagna}, {Spoto}, {Stampa}, {Steele}, {Steidelm{\"u}ller}, {Stephenson}, {Stoev}, {Suess}, {S{\"u}veges}, {Surdej}, {Szabados}, {Szegedi-Elek}, {Tapiador}, {Taris}, {Tauran}, {Taylor}, {Teixeira}, {Terrett}, {Tingley}, {Trager}, {Turon}, {Ulla}, {Utrilla}, {Valentini}, {van Elteren}, {Van Hemelryck}, {van Leeuwen}, {Varadi}, {Vecchiato}, {Veljanoski}, {Via}, {Vicente}, {Vogt}, {Voss}, {Votruba}, {Voutsinas}, {Walmsley}, {Weiler}, {Weingrill}, {Werner}, {Wevers}, {Whitehead}, {Wyrzykowski}, {Yoldas}, {{\v{Z}}erjal}, {Zucker}, {Zurbach}, {Zwitter}, {Alecu}, {Allen}, {Allende Prieto}, {Amorim}, {Anglada-Escud{\'e}}, {Arsenijevic}, {Azaz}, {Balm}, {Beck},
  {Bernstein}, {Bigot}, {Bijaoui}, {Blasco}, {Bonfigli}, {Bono}, {Boudreault}, {Bressan}, {Brown}, {Brunet}, {Bunclark}, {Buonanno}, {Butkevich}, {Carret}, {Carrion}, {Chemin}, {Ch{\'e}reau}, {Corcione}, {Darmigny}, {de Boer}, {de Teodoro}, {de Zeeuw}, {Delle Luche}, {Domingues}, {Dubath}, {Fodor}, {Fr{\'e}zouls}, {Fries}, {Fustes}, {Fyfe}, {Gallardo}, {Gallegos}, {Gardiol}, {Gebran}, {Gomboc}, {G{\'o}mez}, {Grux}, {Gueguen}, {Heyrovsky}, {Hoar}, {Iannicola}, {Isasi Parache}, {Janotto}, {Joliet}, {Jonckheere}, {Keil}, {Kim}, {Klagyivik}, {Klar}, {Knude}, {Kochukhov}, {Kolka}, {Kos}, {Kutka}, {Lainey}, {LeBouquin}, {Liu}, {Loreggia}, {Makarov}, {Marseille}, {Martayan}, {Martinez-Rubi}, {Massart}, {Meynadier}, {Mignot}, {Munari}, {Nguyen}, {Nordlander}, {Ocvirk}, {O'Flaherty}, {Olias Sanz}, {Ortiz}, {Osorio}, {Oszkiewicz}, {Ouzounis}, {Palmer}, {Park}, {Pasquato}, {Peltzer}, {Peralta}, {P{\'e}turaud}, {Pieniluoma}, {Pigozzi}, {Poels}, {Prat}, {Prod'homme}, {Raison}, {Rebordao}, {Risquez}, {Rocca-Volmerange},
  {Rosen}, {Ruiz-Fuertes}, {Russo}, {Sembay}, {Serraller Vizcaino}, {Short}, {Siebert}, {Silva}, {Sinachopoulos}, {Slezak}, {Soffel}, {Sosnowska}, {Strai{\v{z}}ys}, {ter Linden}, {Terrell}, {Theil}, {Tiede}, {Troisi}, {Tsalmantza}, {Tur}, {Vaccari}, {Vachier}, {Valles}, {Van Hamme}, {Veltz}, {Virtanen}, {Wallut}, {Wichmann}, {Wilkinson}, {Ziaeepour}, \& {Zschocke}}]{2016A&A...595A...1G}
{Gaia Collaboration}, {Prusti}, T., {de Bruijne}, J.~H.~J., {et~al.} 2016, \bibinfo{title}{{The Gaia mission},} \aap, 595, A1, \dodoi{10.1051/0004-6361/201629272}

\bibitem[{ {Gaia Collaboration} {et~al.}(2018){Gaia Collaboration}, {Helmi}, {van Leeuwen}, {McMillan}, {Massari}, {Antoja}, {Robin}, {Lindegren}, {Bastian}, {Arenou}, {Babusiaux}, {Biermann}, {Breddels}, {Hobbs}, {Jordi}, {Pancino}, {Reyl{\'e}}, {Veljanoski}, {Brown}, {Vallenari}, {Prusti}, {de Bruijne}, {Bailer-Jones}, {Evans}, {Eyer}, {Jansen}, {Klioner}, {Lammers}, {Luri}, {Mignard}, {Panem}, {Pourbaix}, {Randich}, {Sartoretti}, {Siddiqui}, {Soubiran}, {Walton}, {Cropper}, {Drimmel}, {Katz}, {Lattanzi}, {Bakker}, {Cacciari}, {Casta{\~n}eda}, {Chaoul}, {Cheek}, {De Angeli}, {Fabricius}, {Guerra}, {Holl}, {Masana}, {Messineo}, {Mowlavi}, {Nienartowicz}, {Panuzzo}, {Portell}, {Riello}, {Seabroke}, {Tanga}, {Th{\'e}venin}, {Gracia-Abril}, {Comoretto}, {Garcia-Reinaldos}, {Teyssier}, {Altmann}, {Andrae}, {Audard}, {Bellas-Velidis}, {Benson}, {Berthier}, {Blomme}, {Burgess}, {Busso}, {Carry}, {Cellino}, {Clementini}, {Clotet}, {Creevey}, {Davidson}, {De Ridder}, {Delchambre}, {Dell'Oro}, {Ducourant},
  {Fern{\'a}ndez-Hern{\'a}ndez}, {Fouesneau}, {Fr{\'e}mat}, {Galluccio}, {Garc{\'\i}a-Torres}, {Gonz{\'a}lez-N{\'u}{\~n}ez}, {Gonz{\'a}lez-Vidal}, {Gosset}, {Guy}, {Halbwachs}, {Hambly}, {Harrison}, {Hern{\'a}ndez}, {Hestroffer}, {Hodgkin}, {Hutton}, {Jasniewicz}, {Jean-Antoine-Piccolo}, {Jordan}, {Korn}, {Krone-Martins}, {Lanzafame}, {Lebzelter}, {L{\"o}ffler}, {Manteiga}, {Marrese}, {Mart{\'\i}n-Fleitas}, {Moitinho}, {Mora}, {Muinonen}, {Osinde}, {Pauwels}, {Petit}, {Recio-Blanco}, {Richards}, {Rimoldini}, {Sarro}, {Siopis}, {Smith}, {Sozzetti}, {S{\"u}veges}, {Torra}, {van Reeven}, {Abbas}, {Abreu Aramburu}, {Accart}, {Aerts}, {Altavilla}, {{\'A}lvarez}, {Alvarez}, {Alves}, {Anderson}, {Andrei}, {Anglada Varela}, {Antiche}, {Arcay}, {Astraatmadja}, {Bach}, {Baker}, {Balaguer-N{\'u}{\~n}ez}, {Balm}, {Barache}, {Barata}, {Barbato}, {Barblan}, {Barklem}, {Barrado}, {Barros}, {Barstow}, {Bartholom{\'e} Mu{\~n}oz}, {Bassilana}, {Becciani}, {Bellazzini}, {Berihuete}, {Bertone}, {Bianchi}, {Bienaym{\'e}},
  {Blanco-Cuaresma}, {Boch}, {Boeche}, {Bombrun}, {Borrachero}, {Bossini}, {Bouquillon}, {Bourda}, {Bragaglia}, {Bramante}, {Bressan}, {Brouillet}, {Br{\"u}semeister}, {Brugaletta}, {Bucciarelli}, {Burlacu}, {Busonero}, {Butkevich}, {Buzzi}, {Caffau}, {Cancelliere}, {Cannizzaro}, {Cantat-Gaudin}, {Carballo}, {Carlucci}, {Carrasco}, {Casamiquela}, {Castellani}, {Castro-Ginard}, {Charlot}, {Chemin}, {Chiavassa}, {Cocozza}, {Costigan}, {Cowell}, {Crifo}, {Crosta}, {Crowley}, {Cuypers}, {Dafonte}, {Damerdji}, {Dapergolas}, {David}, {David}, {de Laverny}, {De Luise}, {De March}, {de Martino}, {de Souza}, {de Torres}, {Debosscher}, {del Pozo}, {Delbo}, {Delgado}, {Delgado}, {Di Matteo}, {Diakite}, {Diener}, {Distefano}, {Dolding}, {Drazinos}, {Dur{\'a}n}, {Edvardsson}, {Enke}, {Eriksson}, {Esquej}, {Eynard Bontemps}, {Fabre}, {Fabrizio}, {Faigler}, {Falc{\~a}o}, {Farr{\`a}s Casas}, {Federici}, {Fedorets}, {Fernique}, {Figueras}, {Filippi}, {Findeisen}, {Fonti}, {Fraile}, {Fraser}, {Fr{\'e}zouls}, {Gai}, {Galleti},
  {Garabato}, {Garc{\'\i}a-Sedano}, {Garofalo}, {Garralda}, {Gavel}, {Gavras}, {Gerssen}, {Geyer}, {Giacobbe}, {Gilmore}, {Girona}, {Giuffrida}, {Glass}, {Gomes}, {Granvik}, {Gueguen}, {Guerrier}, {Guiraud}, {Guti{\'e}rrez-S{\'a}nchez}, {Hofmann}, {Holland}, {Huckle}, {Hypki}, {Icardi}, {Jan{\ss}en}, {Jevardat de Fombelle}, {Jonker}, {Juh{\'a}sz}, {Julbe}, {Karampelas}, {Kewley}, {Klar}, {Kochoska}, {Kohley}, {Kolenberg}, {Kontizas}, {Kontizas}, {Koposov}, {Kordopatis}, {Kostrzewa-Rutkowska}, {Koubsky}, {Lambert}, {Lanza}, {Lasne}, {Lavigne}, {Le Fustec}, {Le Poncin-Lafitte}, {Lebreton}, {Leccia}, {Leclerc}, {Lecoeur-Taibi}, {Lenhardt}, {Leroux}, {Liao}, {Licata}, {Lindstr{\o}m}, {Lister}, {Livanou}, {Lobel}, {L{\'o}pez}, {Managau}, {Mann}, {Mantelet}, {Marchal}, {Marchant}, {Marconi}, {Marinoni}, {Marschalk{\'o}}, {Marshall}, {Martino}, {Marton}, {Mary}, {Matijevi{\v{c}}}, {Mazeh}, {Messina}, {Michalik}, {Millar}, {Molina}, {Molinaro}, {Moln{\'a}r}, {Montegriffo}, {Mor}, {Morbidelli}, {Morel}, {Morris},
  {Mulone}, {Muraveva}, {Musella}, {Nelemans}, {Nicastro}, {Noval}, {O'Mullane}, {Ord{\'e}novic}, {Ord{\'o}{\~n}ez-Blanco}, {Osborne}, {Pagani}, {Pagano}, {Pailler}, {Palacin}, {Palaversa}, {Panahi}, {Pawlak}, {Piersimoni}, {Pineau}, {Plachy}, {Plum}, {Poggio}, {Poujoulet}, {Pr{\v{s}}a}, {Pulone}, {Racero}, {Ragaini}, {Rambaux}, {Ramos-Lerate}, {Regibo}, {Riclet}, {Ripepi}, {Riva}, {Rivard}, {Rixon}, {Roegiers}, {Roelens}, {Romero-G{\'o}mez}, {Rowell}, {Royer}, {Ruiz-Dern}, {Sadowski}, {Sagrist{\`a} Sell{\'e}s}, {Sahlmann}, {Salgado}, {Salguero}, {Sanna}, {Santana-Ros}, {Sarasso}, {Savietto}, {Schultheis}, {Sciacca}, {Segol}, {Segovia}, {S{\'e}gransan}, {Shih}, {Siltala}, {Silva}, {Smart}, {Smith}, {Solano}, {Solitro}, {Sordo}, {Soria Nieto}, {Souchay}, {Spagna}, {Spoto}, {Stampa}, {Steele}, {Steidelm{\"u}ller}, {Stephenson}, {Stoev}, {Suess}, {Surdej}, {Szabados}, {Szegedi-Elek}, {Tapiador}, {Taris}, {Tauran}, {Taylor}, {Teixeira}, {Terrett}, {Teyssandier}, {Thuillot}, {Titarenko}, {Torra Clotet}, {Turon},
  {Ulla}, {Utrilla}, {Uzzi}, {Vaillant}, {Valentini}, {Valette}, {van Elteren}, {Van Hemelryck}, {van Leeuwen}, {Vaschetto}, {Vecchiato}, {Viala}, {Vicente}, {Vogt}, {von Essen}, {Voss}, {Votruba}, {Voutsinas}, {Walmsley}, {Weiler}, {Wertz}, {Wevems}, {Wyrzykowski}, {Yoldas}, {{\v{Z}}erjal}, {Ziaeepour}, {Zorec}, {Zschocke}, {Zucker}, {Zurbach}, \& {Zwitter}}]{2018A&A...616A..12G}
{Gaia Collaboration}, {Helmi}, A., {van Leeuwen}, F., {et~al.} 2018, \bibinfo{title}{{Gaia Data Release 2. Kinematics of globular clusters and dwarf galaxies around the Milky Way},} \aap, 616, A12, \dodoi{10.1051/0004-6361/201832698}

\bibitem[{ {Gaia Collaboration} {et~al.}(2023){Gaia Collaboration}, {Vallenari}, {Brown}, {Prusti}, {de Bruijne}, {Arenou}, {Babusiaux}, {Biermann}, {Creevey}, {Ducourant}, {Evans}, {Eyer}, {Guerra}, {Hutton}, {Jordi}, {Klioner}, {Lammers}, {Lindegren}, {Luri}, {Mignard}, {Panem}, {Pourbaix}, {Randich}, {Sartoretti}, {Soubiran}, {Tanga}, {Walton}, {Bailer-Jones}, {Bastian}, {Drimmel}, {Jansen}, {Katz}, {Lattanzi}, {van Leeuwen}, {Bakker}, {Cacciari}, {Casta{\~n}eda}, {De Angeli}, {Fabricius}, {Fouesneau}, {Fr{\'e}mat}, {Galluccio}, {Guerrier}, {Heiter}, {Masana}, {Messineo}, {Mowlavi}, {Nicolas}, {Nienartowicz}, {Pailler}, {Panuzzo}, {Riclet}, {Roux}, {Seabroke}, {Sordo}, {Th{\'e}venin}, {Gracia-Abril}, {Portell}, {Teyssier}, {Altmann}, {Andrae}, {Audard}, {Bellas-Velidis}, {Benson}, {Berthier}, {Blomme}, {Burgess}, {Busonero}, {Busso}, {C{\'a}novas}, {Carry}, {Cellino}, {Cheek}, {Clementini}, {Damerdji}, {Davidson}, {de Teodoro}, {Nu{\~n}ez Campos}, {Delchambre}, {Dell'Oro}, {Esquej},
  {Fern{\'a}ndez-Hern{\'a}ndez}, {Fraile}, {Garabato}, {Garc{\'\i}a-Lario}, {Gosset}, {Haigron}, {Halbwachs}, {Hambly}, {Harrison}, {Hern{\'a}ndez}, {Hestroffer}, {Hodgkin}, {Holl}, {Jan{\ss}en}, {Jevardat de Fombelle}, {Jordan}, {Krone-Martins}, {Lanzafame}, {L{\"o}ffler}, {Marchal}, {Marrese}, {Moitinho}, {Muinonen}, {Osborne}, {Pancino}, {Pauwels}, {Recio-Blanco}, {Reyl{\'e}}, {Riello}, {Rimoldini}, {Roegiers}, {Rybizki}, {Sarro}, {Siopis}, {Smith}, {Sozzetti}, {Utrilla}, {van Leeuwen}, {Abbas}, {{\'A}brah{\'a}m}, {Abreu Aramburu}, {Aerts}, {Aguado}, {Ajaj}, {Aldea-Montero}, {Altavilla}, {{\'A}lvarez}, {Alves}, {Anders}, {Anderson}, {Anglada Varela}, {Antoja}, {Baines}, {Baker}, {Balaguer-N{\'u}{\~n}ez}, {Balbinot}, {Balog}, {Barache}, {Barbato}, {Barros}, {Barstow}, {Bartolom{\'e}}, {Bassilana}, {Bauchet}, {Becciani}, {Bellazzini}, {Berihuete}, {Bernet}, {Bertone}, {Bianchi}, {Binnenfeld}, {Blanco-Cuaresma}, {Blazere}, {Boch}, {Bombrun}, {Bossini}, {Bouquillon}, {Bragaglia}, {Bramante}, {Breedt},
  {Bressan}, {Brouillet}, {Brugaletta}, {Bucciarelli}, {Burlacu}, {Butkevich}, {Buzzi}, {Caffau}, {Cancelliere}, {Cantat-Gaudin}, {Carballo}, {Carlucci}, {Carnerero}, {Carrasco}, {Casamiquela}, {Castellani}, {Castro-Ginard}, {Chaoul}, {Charlot}, {Chemin}, {Chiaramida}, {Chiavassa}, {Chornay}, {Comoretto}, {Contursi}, {Cooper}, {Cornez}, {Cowell}, {Crifo}, {Cropper}, {Crosta}, {Crowley}, {Dafonte}, {Dapergolas}, {David}, {David}, {de Laverny}, {De Luise}, {De March}, {De Ridder}, {de Souza}, {de Torres}, {del Peloso}, {del Pozo}, {Delbo}, {Delgado}, {Delisle}, {Demouchy}, {Dharmawardena}, {Di Matteo}, {Diakite}, {Diener}, {Distefano}, {Dolding}, {Edvardsson}, {Enke}, {Fabre}, {Fabrizio}, {Faigler}, {Fedorets}, {Fernique}, {Fienga}, {Figueras}, {Fournier}, {Fouron}, {Fragkoudi}, {Gai}, {Garcia-Gutierrez}, {Garcia-Reinaldos}, {Garc{\'\i}a-Torres}, {Garofalo}, {Gavel}, {Gavras}, {Gerlach}, {Geyer}, {Giacobbe}, {Gilmore}, {Girona}, {Giuffrida}, {Gomel}, {Gomez}, {Gonz{\'a}lez-N{\'u}{\~n}ez},
  {Gonz{\'a}lez-Santamar{\'\i}a}, {Gonz{\'a}lez-Vidal}, {Granvik}, {Guillout}, {Guiraud}, {Guti{\'e}rrez-S{\'a}nchez}, {Guy}, {Hatzidimitriou}, {Hauser}, {Haywood}, {Helmer}, {Helmi}, {Sarmiento}, {Hidalgo}, {Hilger}, {H{\l}adczuk}, {Hobbs}, {Holland}, {Huckle}, {Jardine}, {Jasniewicz}, {Jean-Antoine Piccolo}, {Jim{\'e}nez-Arranz}, {Jorissen}, {Juaristi Campillo}, {Julbe}, {Karbevska}, {Kervella}, {Khanna}, {Kontizas}, {Kordopatis}, {Korn}, {K{\'o}sp{\'a}l}, {Kostrzewa-Rutkowska}, {Kruszy{\'n}ska}, {Kun}, {Laizeau}, {Lambert}, {Lanza}, {Lasne}, {Le Campion}, {Lebreton}, {Lebzelter}, {Leccia}, {Leclerc}, {Lecoeur-Taibi}, {Liao}, {Licata}, {Lindstr{\o}m}, {Lister}, {Livanou}, {Lobel}, {Lorca}, {Loup}, {Madrero Pardo}, {Magdaleno Romeo}, {Managau}, {Mann}, {Manteiga}, {Marchant}, {Marconi}, {Marcos}, {Marcos Santos}, {Mar{\'\i}n Pina}, {Marinoni}, {Marocco}, {Marshall}, {Martin Polo}, {Mart{\'\i}n-Fleitas}, {Marton}, {Mary}, {Masip}, {Massari}, {Mastrobuono-Battisti}, {Mazeh}, {McMillan}, {Messina}, {Michalik},
  {Millar}, {Mints}, {Molina}, {Molinaro}, {Moln{\'a}r}, {Monari}, {Mongui{\'o}}, {Montegriffo}, {Montero}, {Mor}, {Mora}, {Morbidelli}, {Morel}, {Morris}, {Muraveva}, {Murphy}, {Musella}, {Nagy}, {Noval}, {Oca{\~n}a}, {Ogden}, {Ordenovic}, {Osinde}, {Pagani}, {Pagano}, {Palaversa}, {Palicio}, {Pallas-Quintela}, {Panahi}, {Payne-Wardenaar}, {Pe{\~n}alosa Esteller}, {Penttil{\"a}}, {Pichon}, {Piersimoni}, {Pineau}, {Plachy}, {Plum}, {Poggio}, {Pr{\v{s}}a}, {Pulone}, {Racero}, {Ragaini}, {Rainer}, {Raiteri}, {Rambaux}, {Ramos}, {Ramos-Lerate}, {Re Fiorentin}, {Regibo}, {Richards}, {Rios Diaz}, {Ripepi}, {Riva}, {Rix}, {Rixon}, {Robichon}, {Robin}, {Robin}, {Roelens}, {Rogues}, {Rohrbasser}, {Romero-G{\'o}mez}, {Rowell}, {Royer}, {Ruz Mieres}, {Rybicki}, {Sadowski}, {S{\'a}ez N{\'u}{\~n}ez}, {Sagrist{\`a} Sell{\'e}s}, {Sahlmann}, {Salguero}, {Samaras}, {Sanchez Gimenez}, {Sanna}, {Santove{\~n}a}, {Sarasso}, {Schultheis}, {Sciacca}, {Segol}, {Segovia}, {S{\'e}gransan}, {Semeux}, {Shahaf}, {Siddiqui}, {Siebert},
  {Siltala}, {Silvelo}, {Slezak}, {Slezak}, {Smart}, {Snaith}, {Solano}, {Solitro}, {Souami}, {Souchay}, {Spagna}, {Spina}, {Spoto}, {Steele}, {Steidelm{\"u}ller}, {Stephenson}, {S{\"u}veges}, {Surdej}, {Szabados}, {Szegedi-Elek}, {Taris}, {Taylor}, {Teixeira}, {Tolomei}, {Tonello}, {Torra}, {Torra}, {Torralba Elipe}, {Trabucchi}, {Tsounis}, {Turon}, {Ulla}, {Unger}, {Vaillant}, {van Dillen}, {van Reeven}, {Vanel}, {Vecchiato}, {Viala}, {Vicente}, {Voutsinas}, {Weiler}, {Wevers}, {Wyrzykowski}, {Yoldas}, {Yvard}, {Zhao}, {Zorec}, {Zucker}, \& {Zwitter}}]{2023A&A...674A...1G}
{Gaia Collaboration}, {Vallenari}, A., {Brown}, A.~G.~A., {et~al.} 2023, \bibinfo{title}{{Gaia Data Release 3. Summary of the content and survey properties},} \aap, 674, A1, \dodoi{10.1051/0004-6361/202243940}

\bibitem[{J.~E. {Gaustad} {et~al.}(2001){Gaustad}, {McCullough}, {Rosing}, \& {Van Buren}}]{2001PASP..113.1326G}
{Gaustad}, J.~E., {McCullough}, P.~R., {Rosing}, W., \& {Van Buren}, D. 2001, \bibinfo{title}{{A Robotic Wide-Angle H{\ensuremath{\alpha}} Survey of the Southern Sky},} \pasp, 113, 1326, \dodoi{10.1086/323969}

\bibitem[{D. {Graczyk} {et~al.}(2014){Graczyk}, {Pietrzy{\'n}ski}, {Thompson}, {Gieren}, {Pilecki}, {Konorski}, {Udalski}, {Soszy{\'n}ski}, {Villanova}, {G{\'o}rski}, {Suchomska}, {Karczmarek}, {Kudritzki}, {Bresolin}, \& {Gallenne}}]{2014ApJ...780...59G}
{Graczyk}, D., {Pietrzy{\'n}ski}, G., {Thompson}, I.~B., {et~al.} 2014, \bibinfo{title}{{The Araucaria Project. The Distance to the Small Magellanic Cloud from Late-type Eclipsing Binaries},} \apj, 780, 59, \dodoi{10.1088/0004-637X/780/1/59}

\bibitem[{J.~V. {Hindman} {et~al.}(1963){Hindman}, {Kerr}, \& {McGee}}]{1963AuJPh..16..570H}
{Hindman}, J.~V., {Kerr}, F.~J., \& {McGee}, R.~X. 1963, \bibinfo{title}{{A Low Resolution Hydrogen-line Survey of the Magellanic System. II. Interpretation of Results},} Australian Journal of Physics, 16, 570, \dodoi{10.1071/PH630570}

\bibitem[{J.~D. {Hunter}(2007){Hunter}}]{2007CSE.....9...90H}
{Hunter}, J.~D. 2007, \bibinfo{title}{{Matplotlib: A 2D Graphics Environment},} Computing in Science and Engineering, 9, 90, \dodoi{10.1109/MCSE.2007.55}

\bibitem[{M.~J. {Irwin} {et~al.}(1985){Irwin}, {Kunkel}, \& {Demers}}]{1985Natur.318..160I}
{Irwin}, M.~J., {Kunkel}, W.~E., \& {Demers}, S. 1985, \bibinfo{title}{{A blue stellar population in the HI bridge between the two Magellanic Clouds},} \nat, 318, 160, \dodoi{10.1038/318160a0}

\bibitem[{A.~M. {Jacyszyn-Dobrzeniecka} {et~al.}(2017){Jacyszyn-Dobrzeniecka}, {Skowron}, {Mr{\'o}z}, {Soszy{\'n}ski}, {Udalski}, {Pietrukowicz}, {Skowron}, {Poleski}, {Koz{\l}owski}, {Wyrzykowski}, {Pawlak}, {Szyma{\'n}ski}, \& {Ulaczyk}}]{2017AcA....67....1J}
{Jacyszyn-Dobrzeniecka}, A.~M., {Skowron}, D.~M., {Mr{\'o}z}, P., {et~al.} 2017, \bibinfo{title}{{OGLE-ing the Magellanic System: Three-Dimensional Structure of the Clouds and the Bridge using RR Lyrae Stars},} \actaa, 67, 1, \dodoi{10.32023/0001-5237/67.1.1}

\bibitem[{P. {Marigo} {et~al.}(2017){Marigo}, {Girardi}, {Bressan}, {Rosenfield}, {Aringer}, {Chen}, {Dussin}, {Nanni}, {Pastorelli}, {Rodrigues}, {Trabucchi}, {Bladh}, {Dalcanton}, {Groenewegen}, {Montalb{\'a}n}, \& {Wood}}]{2017ApJ...835...77M}
{Marigo}, P., {Girardi}, L., {Bressan}, A., {et~al.} 2017, \bibinfo{title}{{A New Generation of PARSEC-COLIBRI Stellar Isochrones Including the TP-AGB Phase},} \apj, 835, 77, \dodoi{10.3847/1538-4357/835/1/77}

\bibitem[{D.~S. {Mathewson} {et~al.}(1974){Mathewson}, {Cleary}, \& {Murray}}]{1974ApJ...190..291M}
{Mathewson}, D.~S., {Cleary}, M.~N., \& {Murray}, J.~D. 1974, \bibinfo{title}{{The Magellanic Stream.},} \apj, 190, 291, \dodoi{10.1086/152875}

\bibitem[{D.~S. {Mathewson} {et~al.}(1988){Mathewson}, {Ford}, \& {Visvanathan}}]{1988ApJ...333..617M}
{Mathewson}, D.~S., {Ford}, V.~L., \& {Visvanathan}, N. 1988, \bibinfo{title}{{The Structure of the Small Magellanic Cloud. II.},} \apj, 333, 617, \dodoi{10.1086/166772}

\bibitem[{N.~M. {McClure-Griffiths} {et~al.}(2018){McClure-Griffiths}, {D{\'e}nes}, {Dickey}, {Stanimirovi{\'c}}, {}, {Staveley-Smith}, {Jameson}, {Di Teodoro}, {Allison}, {Collier}, {Chippendale}, {Franzen}, {G{\"u}rkan}, {Heald}, {Hotan}, {Kleiner}, {Lee-Waddell}, {McConnell}, {Popping}, {Rhee}, {Riseley}, {Voronkov}, \& {Whiting}}]{2018NatAs...2..901M}
{McClure-Griffiths}, N.~M., {D{\'e}nes}, H., {Dickey}, J.~M., {et~al.} 2018, \bibinfo{title}{{Cold gas outflows from the Small Magellanic Cloud traced with ASKAP},} Nature Astronomy, 2, 901, \dodoi{10.1038/s41550-018-0608-8}

\bibitem[{J.~P. {McMullin} {et~al.}(2007){McMullin}, {Waters}, {Schiebel}, {Young}, \& {Golap}}]{2007ASPC..376..127M}
{McMullin}, J.~P., {Waters}, B., {Schiebel}, D., {Young}, W., \& {Golap}, K. 2007, in Astronomical Society of the Pacific Conference Series, Vol. 376, Astronomical Data Analysis Software and Systems XVI, ed. R.~A. {Shaw}, F.~{Hill}, \& D.~J. {Bell}, 127

\bibitem[{E. {Muller} \& K. {Bekki}(2007){Muller} \& {Bekki}}]{2007MNRAS.381L..11M}
{Muller}, E., \& {Bekki}, K. 2007, \bibinfo{title}{{The origin of large-scale HI structures in the Magellanic Bridge},} \mnras, 381, L11, \dodoi{10.1111/j.1745-3933.2007.00356.x}

\bibitem[{C.~E. {Murray} {et~al.}(2019){Murray}, {Peek}, {Di Teodoro}, {McClure-Griffiths}, {Dickey}, \& {D{\'e}nes}}]{2019ApJ...887..267M}
{Murray}, C.~E., {Peek}, J.~E.~G., {Di Teodoro}, E.~M., {et~al.} 2019, \bibinfo{title}{{The 3D Kinematics of Gas in the Small Magellanic Cloud},} \apj, 887, 267, \dodoi{10.3847/1538-4357/ab510f}

\bibitem[{S. {Nakano} {et~al.}(2025){Nakano}, {Tachihara}, \& {Tamashiro}}]{2025arXiv250212251N}
{Nakano}, S., {Tachihara}, K., \& {Tamashiro}, M. 2025, \bibinfo{title}{{Evidence of Galactic Interaction in the Small Magellanic Cloud Probed by Gaia Selected Massive Star Candidates},} arXiv e-prints, arXiv:2502.12251, \dodoi{10.48550/arXiv.2502.12251}

\bibitem[{F. {Niederhofer} {et~al.}(2021){Niederhofer}, {Cioni}, {Rubele}, {Schmidt}, {Diaz}, {Matijev{\u{i}}c}, {Bekki}, {Bell}, {de Grijs}, {El Youssoufi}, {Ivanov}, {Oliveira}, {Ripepi}, {Subramanian}, {Sun}, \& {van Loon}}]{2021MNRAS.502.2859N}
{Niederhofer}, F., {Cioni}, M.-R.~L., {Rubele}, S., {et~al.} 2021, \bibinfo{title}{{The VMC survey - XLI. Stellar proper motions within the Small Magellanic Cloud},} \mnras, 502, 2859, \dodoi{10.1093/mnras/stab206}

\bibitem[{F. Ochsenbein(1996)Ochsenbein}]{10.26093/cds/vizier}
Ochsenbein, F. 1996, \bibinfo{title}{The VizieR database of astronomical catalogues,} CDS, Centre de Données astronomiques de Strasbourg, \dodoi{10.26093/CDS/VIZIER}

\bibitem[{F. {Ochsenbein} {et~al.}(2000){Ochsenbein}, {Bauer}, \& {Marcout}}]{vizier2000}
{Ochsenbein}, F., {Bauer}, P., \& {Marcout}, J. 2000, \bibinfo{title}{{The VizieR database of astronomical catalogues},} \aaps, 143, 23, \dodoi{10.1051/aas:2000169}

\bibitem[{M.~S. {Oey} {et~al.}(2018){Oey}, {Dorigo Jones}, {Castro}, {Zivick}, {Besla}, {Januszewski}, {Moe}, {Kallivayalil}, \& {Lennon}}]{2018ApJ...867L...8O}
{Oey}, M.~S., {Dorigo Jones}, J., {Castro}, N., {et~al.} 2018, \bibinfo{title}{{Resolved Kinematics of Runaway and Field OB Stars in the Small Magellanic Cloud},} \apjl, 867, L8, \dodoi{10.3847/2041-8213/aae892}

\bibitem[{ {OpenAI} {et~al.}(2024){OpenAI}, {:}, {Hurst}, {Lerer}, {Goucher}, {Perelman}, {Ramesh}, {Clark}, {Ostrow}, {Welihinda}, {Hayes}, {Radford}, {M{\k{a}}dry}, {Baker-Whitcomb}, {Beutel}, {Borzunov}, {Carney}, {Chow}, {Kirillov}, {Nichol}, {Paino}, {Renzin}, {Tachard Passos}, {Kirillov}, {Christakis}, {Conneau}, {Kamali}, {Jabri}, {Moyer}, {Tam}, {Crookes}, {Tootoochian}, {Tootoonchian}, {Kumar}, {Vallone}, {Karpathy}, {Braunstein}, {Cann}, {Codispoti}, {Galu}, {Kondrich}, {Tulloch}, {Mishchenko}, {Baek}, {Jiang}, {Pelisse}, {Woodford}, {Gosalia}, {Dhar}, {Pantuliano}, {Nayak}, {Oliver}, {Zoph}, {Ghorbani}, {Leimberger}, {Rossen}, {Sokolowsky}, {Wang}, {Zweig}, {Hoover}, {Samic}, {McGrew}, {Spero}, {Giertler}, {Cheng}, {Lightcap}, {Walkin}, {Quinn}, {Guarraci}, {Hsu}, {Kellogg}, {Eastman}, {Lugaresi}, {Wainwright}, {Bassin}, {Hudson}, {Chu}, {Nelson}, {Li}, {Shern}, {Conger}, {Barette}, {Voss}, {Ding}, {Lu}, {Zhang}, {Beaumont}, {Hallacy}, {Koch}, {Gibson}, {Kim}, {Choi}, {McLeavey}, {Hesse},
  {Fischer}, {Winter}, {Czarnecki}, {Jarvis}, {Wei}, {Koumouzelis}, {Sherburn}, {Kappler}, {Levin}, {Levy}, {Carr}, {Farhi}, {Mely}, {Robinson}, {Sasaki}, {Jin}, {Valladares}, {Tsipras}, {Li}, {Nguyen}, {Findlay}, {Oiwoh}, {Wong}, {Asdar}, {Proehl}, {Yang}, {Antonow}, {Kramer}, {Peterson}, {Sigler}, {Wallace}, {Brevdo}, {Mays}, {Khorasani}, {Petroski Such}, {Raso}, {Zhang}, {von Lohmann}, {Sulit}, {Goh}, {Oden}, {Salmon}, {Starace}, {Brockman}, {Salman}, {Bao}, {Hu}, {Wong}, {Wang}, {Schmidt}, {Whitney}, {Jun}, {Kirchner}, {Ponde de Oliveira Pinto}, {Ren}, {Chang}, {Chung}, {Kivlichan}, {O'Connell}, {O'Connell}, {Osband}, {Silber}, {Sohl}, {Okuyucu}, {Lan}, {Kostrikov}, {Sutskever}, {Kanitscheider}, {Gulrajani}, {Coxon}, {Menick}, {Pachocki}, {Aung}, {Betker}, {Crooks}, {Lennon}, {Kiros}, {Leike}, {Park}, {Kwon}, {Phang}, {Teplitz}, {Wei}, {Wolfe}, {Chen}, {Harris}, {Varavva}, {Lee}, {Shieh}, {Lin}, {Yu}, {Weng}, {Tang}, {Yu}, {Jang}, {Quinonero Candela}, {Beutler}, {Landers}, {Parish}, {Heidecke},
  {Schulman}, {Lachman}, {McKay}, {Uesato}, {Ward}, \& {Kim}}]{2024arXiv241021276O}
{OpenAI}, {:}, {Hurst}, A., {et~al.} 2024, \bibinfo{title}{{GPT-4o System Card},} arXiv e-prints, arXiv:2410.21276, \dodoi{10.48550/arXiv.2410.21276}

\bibitem[{G. {Pastorelli} {et~al.}(2019){Pastorelli}, {Marigo}, {Girardi}, {Chen}, {Rubele}, {Trabucchi}, {Aringer}, {Bladh}, {Bressan}, {Montalb{\'a}n}, {Boyer}, {Dalcanton}, {Eriksson}, {Groenewegen}, {H{\"o}fner}, {Lebzelter}, {Nanni}, {Rosenfield}, {Wood}, \& {Cioni}}]{2019MNRAS.485.5666P}
{Pastorelli}, G., {Marigo}, P., {Girardi}, L., {et~al.} 2019, \bibinfo{title}{{Constraining the thermally pulsing asymptotic giant branch phase with resolved stellar populations in the Small Magellanic Cloud},} \mnras, 485, 5666, \dodoi{10.1093/mnras/stz725}

\bibitem[{G. {Pastorelli} {et~al.}(2020){Pastorelli}, {Marigo}, {Girardi}, {Aringer}, {Chen}, {Rubele}, {Trabucchi}, {Bladh}, {Boyer}, {Bressan}, {Dalcanton}, {Groenewegen}, {Lebzelter}, {Mowlavi}, {Chubb}, {Cioni}, {de Grijs}, {Ivanov}, {Nanni}, {van Loon}, \& {Zaggia}}]{2020MNRAS.498.3283P}
{Pastorelli}, G., {Marigo}, P., {Girardi}, L., {et~al.} 2020, \bibinfo{title}{{Constraining the thermally pulsing asymptotic giant branch phase with resolved stellar populations in the Large Magellanic Cloud},} \mnras, 498, 3283, \dodoi{10.1093/mnras/staa2565}

\bibitem[{V. {Ripepi} {et~al.}(2016){Ripepi}, {Marconi}, {Moretti}, {Clementini}, {Cioni}, {de Grijs}, {Emerson}, {Groenewegen}, {Ivanov}, \& {Piatti}}]{2016ApJS..224...21R}
{Ripepi}, V., {Marconi}, M., {Moretti}, M.~I., {et~al.} 2016, \bibinfo{title}{{The VMC Survey. XIX. Classical Cepheids in the Small Magellanic Cloud},} \apjs, 224, 21, \dodoi{10.3847/0067-0049/224/2/21}

\bibitem[{V. {Ripepi} {et~al.}(2017){Ripepi}, {Cioni}, {Moretti}, {Marconi}, {Bekki}, {Clementini}, {de Grijs}, {Emerson}, {Groenewegen}, {Ivanov}, {Molinaro}, {Muraveva}, {Oliveira}, {Piatti}, {Subramanian}, \& {van Loon}}]{2017MNRAS.472..808R}
{Ripepi}, V., {Cioni}, M.-R.~L., {Moretti}, M.~I., {et~al.} 2017, \bibinfo{title}{{The VMC survey - XXV. The 3D structure of the Small Magellanic Cloud from Classical Cepheids},} \mnras, 472, 808, \dodoi{10.1093/mnras/stx2096}

\bibitem[{T. {Robitaille} \& E. {Bressert}(2012){Robitaille} \& {Bressert}}]{2012ascl.soft08017R}
{Robitaille}, T., \& {Bressert}, E. 2012, \bibinfo{title}{{APLpy: Astronomical Plotting Library in Python},}, Astrophysics Source Code Library, record ascl:1208.017

\bibitem[{M. {Salem} {et~al.}(2015){Salem}, {Besla}, {Bryan}, {Putman}, {van der Marel}, \& {Tonnesen}}]{2015ApJ...815...77S}
{Salem}, M., {Besla}, G., {Bryan}, G., {et~al.} 2015, \bibinfo{title}{{Ram Pressure Stripping of the Large Magellanic Cloud's Disk as a Probe of the Milky Way's Circumgalactic Medium},} \apj, 815, 77, \dodoi{10.1088/0004-637X/815/1/77}

\bibitem[{I. {Soszy{\'n}ski} {et~al.}(2015){Soszy{\'n}ski}, {Udalski}, {Szyma{\'n}ski}, {Skowron}, {Pietrzy{\'n}ski}, {Poleski}, {Pietrukowicz}, {Skowron}, {Mr{\'o}z}, {Koz{\l}owski}, {Wyrzykowski}, {Ulaczyk}, \& {Pawlak}}]{2015AcA....65..297S}
{Soszy{\'n}ski}, I., {Udalski}, A., {Szyma{\'n}ski}, M.~K., {et~al.} 2015, \bibinfo{title}{{The OGLE Collection of Variable Stars. Classical Cepheids in the Magellanic System},} \actaa, 65, 297, \dodoi{10.48550/arXiv.1601.01318}

\bibitem[{S. {Stanimirovi{\'c}} {et~al.}(2004){Stanimirovi{\'c}}, {Staveley-Smith}, \& {Jones}}]{2004ApJ...604..176S}
{Stanimirovi{\'c}}, S., {Staveley-Smith}, L., \& {Jones}, P.~A. 2004, \bibinfo{title}{{A New Look at the Kinematics of Neutral Hydrogen in the Small Magellanic Cloud},} \apj, 604, 176, \dodoi{10.1086/381869}

\bibitem[{J. {Tang} {et~al.}(2014){Tang}, {Bressan}, {Rosenfield}, {Slemer}, {Marigo}, {Girardi}, \& {Bianchi}}]{2014MNRAS.445.4287T}
{Tang}, J., {Bressan}, A., {Rosenfield}, P., {et~al.} 2014, \bibinfo{title}{{New PARSEC evolutionary tracks of massive stars at low metallicity: testing canonical stellar evolution in nearby star-forming dwarf galaxies},} \mnras, 445, 4287, \dodoi{10.1093/mnras/stu2029}

\bibitem[{A. {Udalski} {et~al.}(2015){Udalski}, {Szyma{\'n}ski}, \& {Szyma{\'n}ski}}]{2015AcA....65....1U}
{Udalski}, A., {Szyma{\'n}ski}, M.~K., \& {Szyma{\'n}ski}, G. 2015, \bibinfo{title}{{OGLE-IV: Fourth Phase of the Optical Gravitational Lensing Experiment},} \actaa, 65, 1, \dodoi{10.48550/arXiv.1504.05966}

\bibitem[{P. {Zivick} {et~al.}(2018){Zivick}, {Kallivayalil}, {van der Marel}, {Besla}, {Linden}, {Koz{\l}owski}, {Fritz}, {Kochanek}, {Anderson}, {Sohn}, {Geha}, \& {Alcock}}]{2018ApJ...864...55Z}
{Zivick}, P., {Kallivayalil}, N., {van der Marel}, R.~P., {et~al.} 2018, \bibinfo{title}{{The Proper Motion Field of the Small Magellanic Cloud: Kinematic Evidence for Its Tidal Disruption},} \apj, 864, 55, \dodoi{10.3847/1538-4357/aad4b0}

\end{thebibliography}
\bibliographystyle{aasjournalv7}



\end{document}